\documentclass[12pt,a4paper,prd,nofootinbib, superscriptaddress]{revtex4-1}

\usepackage{graphicx}
\usepackage{amsmath,amsthm}
\usepackage{amssymb,amsfonts}
\usepackage{hyperref}
\usepackage{slashed} 
\usepackage{diagbox} 
\usepackage{eucal}
\usepackage{extarrows}
\usepackage{mathtools} 
\usepackage{tikz}
\usepackage{tikz-cd}
\usetikzlibrary{arrows.meta} 
\usetikzlibrary{calc}
\usetikzlibrary{shapes.geometric,calc}
\usetikzlibrary{matrix,arrows,decorations.pathmorphing}

\newlength{\hght}
\newlength{\wdth}
\newcommand{\cancelto}[3][.5ex]{\setlength{\hght}{\heightof{$#3$}}\setlength{\wdth}{\widthof{$#3$}}%
\makebox[0pt][l]{\tikz[baseline]{\draw[-latex](0,-#1)--(\wdth,\hght+#1)
node[shift={(1mm,.5mm)}]{$#2$};}}#3}
\newcommand{\xcancelto}[3][.5ex]{\setlength{\hght}{\heightof{$#3$}}\setlength{\wdth}{\widthof{$#3$}}%
    \makebox[0pt][l]{\tikz[baseline]{\draw[latex-latex](0,-#1)--(\wdth,\hght+#1)
    node[shift={(1mm,.5mm)}]{$#2$};}}#3}

\newcommand{\circled}[1]{%
  \tikz[baseline=(char.base)]\node[anchor=south west, draw = blue,rectangle,
    rounded corners, inner sep=2pt, minimum size=4mm, text
    height=1mm](char){\ensuremath{#1}} ;} 
\newcommand{\greencircled}[1]{%
\tikz[baseline=(char.base)]\node[anchor=south west, draw = green,rectangle, 
    inner sep=2pt, minimum size=4mm, text
    height=1mm](char){\ensuremath{#1}}
     ;}
\usepackage[margin=2.5cm]{geometry}

\newcommand{\R}{\mathbb{R}} 
\newcommand{\C}{\mathbb{C}}
 
\newcommand{\Z}{\mathbb{Z}}
\newcommand{\cA}{\mathcal{A}} 

\newcommand{\cD}{\mathcal{D}}

\newcommand{\cO}{\mathcal{O}}

\definecolor{mygreen}{rgb}{0.13,0.55,0.13}

\newcommand{\Q}{{Q}} 
\def\delbar{\bar\partial}

\DeclareMathOperator{\Sym}{Sym}

\numberwithin{subsubsection}{subsection}
\numberwithin{subsection}{section}
\numberwithin{equation}{section}

\makeatletter
\def\p@subsection{} \def\p@subsubsection{}

\newcommand{\nocontentsline}[3]{}
\newcommand{\tocless}[2]{\bgroup\let\addcontentsline=\nocontentsline#1{#2}\egroup}

\renewcommand\section{\@startsection {section}{1}{\z@}%
                                   {-3.5ex \@plus -1ex \@minus -.2ex}%
                                   {2.3ex \@plus.2ex}%
                                   {\normalfont\large\bfseries}}
\renewcommand\subsection{\@startsection{subsection}{2}{\z@}%
                                   {-3.25ex\@plus -1ex \@minus -.2ex}%
                                   {1.5ex \@plus .2ex}%
                                   {\normalfont\normalsize\bfseries}}
\renewcommand\subsubsection{\@startsection{subsubsection}{3}{\z@}%
                                   {-3.25ex\@plus -1ex \@minus -.2ex}%
                                   {1.5ex \@plus .2ex}%
                                   {\normalfont\normalsize\it}}
\renewcommand\paragraph{\@startsection{paragraph}{4}{\z@}%
                                   {-3.25ex\@plus -1ex \@minus -.2ex}%
                                   {1.5ex \@plus .2ex}%
                                   {\normalfont\normalsize\bf}}
\renewcommand\subparagraph{\@startsection{subparagraph}{5}{\z@}%
                                   {-1.25ex\@plus -1ex \@minus -.2ex}%
                                   {1.5ex \@plus .2ex}%
                                   {\normalfont\normalsize\it}}

\begin{document}

\title{Maximally twisted eleven-dimensional supergravity}

\author{Richard Eager}
\affiliation{Kishine Koen \\ 222-0034 Yokohama, Japan}
\email{eager@mathi.uni-heidelberg.de}
\author{Fabian Hahner}
\affiliation{Mathematisches Institut der Universit\"at Heidelberg \\ Im
Neuenheimer Feld 205 \\ 69120 Heidelberg, Germany}
\email{fhahner@mathi.uni-heidelberg.de}

\begin{abstract}
We perform the maximal twist of eleven-dimensional supergravity. This twist is
partially topological and exists on manifolds of $G_2 \times SU(2)$ holonomy.
Our derivation starts with an explicit description of the Batalin-Vilkovisky
complex associated to the three-form multiplet in the pure spinor superfield
formalism. We then determine the $L_\infty$ module structure of the
supersymmetry algebra on the component fields. We twist the theory by modifying
the differential of the Batalin-Vilkovisky complex to incorporate the action of
a scalar supercharge. We find that the resulting free twisted theory is given by
the tensor product of the de Rham and Dolbeault complexes of the respective
$G_2$ and $SU(2)$ holonomy manifolds as conjectured by Costello.
\end{abstract}

\maketitle

\addtocounter{page}{-1}

\setcounter{tocdepth}{2}
\tableofcontents

\tocless\bibsection

\newpage 
\hfill\parbox[b]{10cm}{{\it In conclusion, two things remain to be done with our
theory.  First, we are studying the reduction to four dimensions...} (Cremmer,
Julia and Scherk \cite{Cremmer:1978km})}

\section{Introduction}
\label{sec:intro}
Eleven-dimensional supergravity \cite{Cremmer:1978km} is the low energy limit of
M-theory, a conjectural theory that is believed to unify type I, II, and
heterotic superstring theories \cite{Witten:1995ex}. It realizes the maximal
dimension that has a supersymmetric representation with particles of spin at
most two \cite{Nahm:1977tg}, and the action of eleven-dimensional supergravity
is unique \cite{Cremmer:1978km}. M-theory compactifications on manifolds with
$G_2$ holonomy result in four-dimensional field theories with minimal
supersymmetry and have been intensely studied in relation to non-perturbative
string dualities and phenomenology.

In this note, we consider a partial topological twist of eleven-dimensional
supergravity on manifolds of $G_2 \times SU(2)$ holonomy. Partial topological
twists are a natural arena where off-shell representations of supersymmetry,
supersymmetric localization, special holonomy manifolds, and elliptic moduli
problems converge. A partial topological twist can only be performed on a
manifold of special holonomy.  The equations of motion after twisting often
simplify to elliptic complexes that are specific to the special holonomy
manifold on which the twist is defined.

A conjectured partial topological twist of eleven-dimensional supergravity on
manifolds, $M^7 \times M^4,$ of $G_2 \times SU(2)$ holonomy is given in
\cite{Costello2015, Costello:2016nkh,Raghavendran:2019zdq}. As a free BV theory,
the twisted theory is described by the cochain complex
\begin{equation}
	\left( \Omega^{\bullet}(M^7) \otimes \Omega^{0,\bullet}(M^4) \: , \: D^{\mathrm{tw}} \right) ,
\end{equation}
where the differential $D$ decomposes into
\begin{equation}
	D^{\mathrm{tw}} = d_{M^7} \otimes 1 + 1 \otimes \delbar_{M^4} \: .
\end{equation}
Here $d_{M^7}$ is the de Rham differential on $M^7$ and $\delbar_{M^4}$ is the
Dolbeault differential on $M^4$. In principle, higher interaction terms will
also be present, but here we restrict our attention to the free theory.  

Already, twisted M-theory has had several applications to mathematical physics
\cite{Costello:2016nkh,  Raghavendran:2019zdq}. While these works are rigorous
mathematics in the sense of Jaffe--Quinn \cite{MR1202292}, our aim is to connect
them to eleven-dimensional supergravity as originally envisioned by
Cremmer--Julia--Scherk \cite{Cremmer:1978km} and its more recent formulations in
the pure spinor formalism \cite{Berkovits:2002uc, Cederwall:2010tn}.  

In this note we will show how to obtain the fields and BV differential by
directly twisting the fields of M-theory in the BV formalism
\cite{Batalin:1981jr}. After the twist, the three-form $C^{(3)}$ with its ghost
system $C^{(2)},C^{(1)},C^{(0)},$ the spin-$3/2$ Rarita--Schwinger field $\psi,$
and all of their corresponding antifields organize into a differential form $\cA
\in \Omega^{\bullet}(M^7) \otimes \Omega^{0,\bullet}(M^4)$, as conjectured by
Costello. Its components are displayed in Table~\ref{tab:twistedfields}.

\begin{table}[htp]
\caption{Fields in $G_2 \times SU(2)$ twisted M-theory}
\begin{center}
\begin{tabular}{|c|cccccccc|}
\hline
& $\Omega^0(M^7)$ & $\Omega^1(M^7)$ & $\Omega^2(M^7)$ & $\Omega^3(M^7)$ &
$\Omega^4(M^7)$ & $\Omega^5(M^7)$ & $\Omega^6(M^7)$ & $\Omega^7(M^7)$ \\
\hline
$\Omega^{0,0}(M^4)$  & $C^{(0)}$ & $C^{(1)}$ & $C^{(2)}$ & $C^{(3)}$ & $\psi$ &
$\psi^{\dagger}$ & $C^{(3)\dagger}$ & $C^{(2)\dagger}$ \\
$\Omega^{0,1}(M^4)$  & $C^{(1)}$ & $C^{(2)}$ & $C^{(3)}$ & $\psi$ &
$\psi^{\dagger}$ & $C^{(3)\dagger}$ & $C^{(2)\dagger}$ & $C^{(1)\dagger}$ \\
$\Omega^{0,2}(M^4)$  & $C^{(2)}$ & $C^{(3)}$ & $\psi$ & $\psi^{\dagger}$ &
$C^{(3) \dagger}$ & $C^{(2)\dagger}$ & $C^{(1)\dagger}$ & $C^{(0)\dagger}$ \\
\hline
\end{tabular}
\end{center}
\label{tab:twistedfields}
\end{table}

We will derive the conjectured form of the twisted fields and differential
starting from the manifestly covariant formulation of eleven-dimensional
supergravity \cite{Berkovits:2002uc, Cederwall:2009ez, Cederwall:2010tn,
Berkovits:2018gbq} in the pure spinor superfield formalism
\cite{Berkovits:2000fe, Berkovits:2005bt, Cederwall:2013vba}. We use this
formalism to obtain the BV complex of the three-form multiplet in eleven
dimensional supergravity, including the full action of the supersymmetry algebra
on the component fields. These results are then used to carry out the actual
twist on the level of component fields. This gives an explicit understanding of
the fields in the twisted theory as well as the formation of trivial pairs in
terms of the fields of the untwisted supergravity multiplet.

The traditional approach to eleven-dimensional supergravity in superspace
\cite{Brink:1980az, Cremmer:1980ru, Cederwall:2000fj,Cederwall:2000ye,
Cederwall:2004cg} starts with the supervielbein and imposes conventional
constraints \cite{Gates:1979wg,Gates:1979jv} on torsions and curvatures.  We
will make some speculative remarks about the twist of the supervielbein at the
end. A partially off-shell formulation of eleven-dimensional supergravity
adapted to manifolds of $G_2 \times SU(2)$ holonomy is given in
\cite{Becker:2017zwe, Becker:2018phr, Becker:2021oiz} and is closely related to
the twisted theory.

We will work in Euclidean signature. We hope to return to the twist of the
higher order terms and the formulation in Lorentzian signature in subsequent
work.

\bigskip
\hfill\parbox[b]{11.5cm}{{\it From the outset you know, more or less, what
became of the three-form multiplet, so most of your curiosity is invested in the
question of how it all came to pass.} (Adapted from A.O. Scott.)}
\subparagraph{Organization:}
The rest of this work is structured as follows.  In Section~\ref{sec:Setup} we
review supersymmetric theories in the BV formalism and explain how to twist them
with respect to a supercharge.  
We describe the types of twists in eleven-dimensional supergravity and the $G_2
\times SU(2)$ invariant twist in detail. In Section~\ref{sec:PS} we specialize
our general discussion of BV theories to eleven-dimensional supergravity. We
introduce the BV complex for the three-form multiplet  and describe the action
of supersymmetry on its component fields. Finally, in
Section~\ref{sec:twistedfields} we describe the decomposition of the fields and
supersymmetry transformations with respect to $G_2 \times SU(2).$ We then use
the decomposition to determine the fields surviving the partial topological
twist and the resulting action of the modified BV differential. We conclude with
some thoughts on further directions in Section~\ref{sec:conclusions}.

\subparagraph{Note added:}
The authors thank Ingmar Saberi and Brian Williams for informing them of their
related paper \cite{sw2021} and coordinating submission to the arXiv.  Their
work derives the holomorphic twist of the eleven-dimensional three-form
multiplet. Starting from the holomorphic twist of Saberi--Williams, Surya
Raghavendran and Brian Williams independently derive the $G_2 \times SU(2)$
invariant twist in their forthcoming work \cite{rw2021}. Our work is
complementary to that of Saberi--Williams and Raghavendran--Williams in the
sense that we determine the origin of the twisted fields in the untwisted
theory, whereas their work cleverly bypasses the component fields of the
untwisted theory. Further discussion of the relations between these different
perspectives will appear in \cite{rfsw2021}.

\section{Twisting \`a la Costello}
\label{sec:Setup}
\subsection{Supersymmetric field theories in the BV formalism} 
\label{sec:SusyandBV}
In the BV formalism, a field theory is described by a sheaf of cyclic (super)
$L_{\infty}$ algebras over a spacetime $M$. This sheaf models the space of
solutions to the equations of motion up to gauge equivalence
\cite{Batalin:1981jr, MR2778558, CostelloGwilliam2021}. Here we are only
concerned with free field theories, which means that all involved $L_\infty$
algebras have no higher operations ($\mu_i=0$ for $i\geq2$) and hence simply are
cochain complexes. Given such an $L_\infty$ algebra $L$, the space $\mathcal{E}$
of BV fields is obtained by a homological shift $\mathcal{E} = L[1]$. The space
of BV fields usually arises as the sections of a $\Z \times \Z/2\Z$ graded
vector bundle $E \rightarrow M$ over the spacetime $M$,
\begin{equation}
	(\mathcal{E}, D) = (\Gamma(M,E), D) \: ,
\end{equation}
where the differential $D$ arises as a differential operator of degree $(1,+)$.
The $\Z$-grading is usually called ghost number, while the $\Z/2\Z$ grading
corresponds to the usual parity distinguishing bosons and fermions. By
assumption, $E$ is equipped with a fiberwise non-degenerate, graded
antisymmetric map of bidegree $(1,+)$
\begin{equation}
	\omega : E \otimes E \longrightarrow \mathrm{Dens}_M,
\end{equation}
which induces a pairing on compactly supported sections $\mathcal{E}_c \subseteq
\mathcal{E}$ via integration. Due to its degree, this pairing connects fields
and antifields.

A free classical BV theory, specified by the data $(E,D,\omega)$, can also be
described in a second way that is, in a sense, dual to the above description.
The pairing $\omega$ endows the functionals $\cO(\mathcal{E})$ of the fields
with an odd Poisson bracket $\{-,-\}$ of degree $1$. The differential $D$
induces a BV operator $Q_{BV} : \cO(\mathcal{E}) \rightarrow \cO(\mathcal{E})$
that can be written in the form
\begin{equation}
	Q_{BV} = \{S_{BV} , -\}
\end{equation}
for the BV action $S_{BV}$, which satisfies the classical master equation
$\{S_{BV}, S_{BV}\} = 0$.  As we are dealing with free theories, the BV action
can be written as
\begin{equation}
	S_{BV}[\Phi] = \int_M \omega(\Phi, D\Phi) \: .
\end{equation}
Thus, for free theories and in the presence of a non-degenerate pairing
$\omega$, the BV operator $Q_{BV}$, the differential $D$ and the BV action
$S_{BV}$ all contain the same information.

An important subset of all functionals of the fields are the local operators:
For any point $x \in M$, we have local operators supported at $x$
\begin{equation}
	\cO_x(\mathcal{E}) = \Sym^\bullet(J^\infty E|_x)^\vee \: ,
\end{equation}
where $J^\infty$ denotes the infinite jet space. For example, given a field
$\phi \in \mathcal{E}$, the corresponding local operator in $\cO_x(\mathcal{E})$
evaluates $\phi$ at the point $x$.

Corresponding to these two point of views, there are also two `dual' ways of
encoding the action of supersymmetry in the BV formalism. Let us start from the
perspective of fields. The endomorphisms $\mathrm{End}(\mathcal{E})$, equipped
with the commutator and the differential $[D,-]$, form a differential graded
super Lie algebra. Inside $(\mathrm{End}(\mathcal{E}), [D,-])$, there is a sub
dg super Lie algebra denoted by $(\cD(\mathcal{E}), [D,-])$, consisting of all
endomorphisms of $\mathcal{E}$ acting by differential operators.

Now let $\mathfrak{p} = \mathfrak{p}_+ \oplus \mathfrak{p}_-$ denote a super
Poincar\'e algebra. In a supersymmetric field theory, the action of
supersymmetry is described on the fields by a map of super $L_\infty$ algebras
\begin{equation}
	\rho: \mathfrak{p} \rightsquigarrow (\cD(\mathcal{E}), [D,-]) \: .
\end{equation}
As such $\rho$ consists of component maps
\begin{equation}
	\rho^{(i)} : \mathfrak{p}^{\otimes i} \longrightarrow \cD(\mathcal{E}) \: , \quad
	 j \geq 1
\end{equation}
of degree $1-i$ satisfying the usual consistency relations for morphisms of
super $L_\infty$ algebras.

The action of supersymmetry can also be encoded on the operators of the theory.
This is done by combining, for $Q \in \mathfrak{p}$,
$\rho^{(i)}(Q,\dots,Q)^\vee$ the dual maps into a differential
\begin{equation}
	\delta_Q = \sum_i \rho^{(i)} (Q,\dots,Q)^\vee : \cO_x(\mathcal{E}) \longrightarrow \cO_x(\mathcal{E}) \: .
\end{equation}
Note that this is the same procedure as encoding a gauge symmetry in the BRST
differential. We will see in Section~\ref{sec:SusyTrafos} how to describe the
action of the supersymmetry algebra in the pure spinor formalism.

\subsection{Twisting in the BV formalism}
Let us fix a square zero element $Q \in \mathfrak{p}_-$ of the odd part of the
supersymmetry algebra. Given a supersymmetric field theory in the BV formalism,
the twist of the theory by $Q$ is defined by deforming the BV operator
\begin{equation}
\label{eq:NewBV}
	Q_{BV} \rightarrow Q_{BV} + t \delta_Q, \: 
\end{equation}
where $t \in \C^{\times}$ and taking $\C^{\times}$-invariants
\cite{Witten:1988ze, Costello:2011np}.  After taking invariants, we specialize
to $t = 1.$ Equivalently, we can deform the BV action to
\begin{equation}
S^Q_{BV}[\Phi] = S_{BV}[\Phi] + \sum_{i} \int_M  \omega(\Phi , \rho^{(i)}(Q, \dots, Q) (\Phi) )
\end{equation}
following \cite{Elliott:2020ecf}. This defines the twisted theory as a classical
BV theory $(E, S^{Q}_{BV}, \omega)$ with the same space of fields and
odd-symplectic pairing, but with a new action. Deforming the differential
typically breaks the grading on $E$. Importantly, the presence of new terms in
the differential often results in the formation of trivial pairs that decouple
and thus play no role in the dynamics of the twisted theory and hence can be
neglected. More precisely, one can pass over to a theory with a smaller space of
fields, where all trivial pairs are omitted. This gives an equivalent
formulation of the twisted theory, which is often drastically simplified. One
usually also refers to this formulation as the twisted theory.

\subsection{Twisting and the Nilpotence Variety}
Eleven-dimensional supergravity can be twisted in two distinct ways that
correspond to the two types of nilpotent supercharges. In this note, we are
exclusively concerned with the maximal twist, which is possible on a manifold
with $G_2 \times SU(2)$ holonomy \cite{Movshev:2011cy, Costello:2016nkh,
Eager:2018dsx}. In addition, there is also a minimal (holomorphic) twist.

The possible twists of a supersymmetric field theory are described by the
variety of square zero elements $Y$ inside the supersymmetry algebra of the
theory. The nilpotence variety
\begin{equation}
	Y = \{ Q \in \mathfrak{p}_- | \{Q,Q\}=0 \}
\end{equation}
has a natural stratification such that each stratum can be identified with a
twisted theory~\cite{Eager:2018oww}. Different strata can be distinguished by
the commutant
\begin{equation}
	Z(Q) = \{ x \in A | [x,Q] = 0 \}, 
\end{equation}
which is constant along the strata.

Recall that, in any dimension, the Dirac spinor representation $S$ is obtained
from a maximal isotropic subspace $L \subset V$ by setting
\begin{equation}
	S = \wedge^\bullet L^\vee \: .
\end{equation}
$S$ forms a Clifford module for $Cl(V)$ and thus in particular a representation
of $\mathfrak{so}(V)$. In the case where $d= \mathrm{dim}(V)$ is odd, this
representation is irreducible. As we are interested in eleven-dimensional
supergravity, we restrict to this case for the moment.

For $Q \in S$, the annihilator with respect to Clifford multiplication
\begin{equation}
	\mathrm{Ann}(Q) = \{ v \in V | v \cdot Q = 0 \}
\end{equation}
gives an isotropic subspace $\mathrm{Ann}(Q) \subset V$. $Q$ is called a
\textit{Cartan pure spinor} if $\mathrm{Ann}(Q)$ is maximal isotropic. Every
Cartan pure spinor is square zero. The converse, however, is in general not true
as we will see below. More generally, one can define the varieties
\begin{equation}
	\mathrm{PS}_k = \{ Q \in S | \mathrm{dim}(L) - \mathrm{dim}(\mathrm{Ann}(Q)) \leq k \} \: ,
\end{equation}
which define a filtration
\begin{equation}
	\mathrm{PS}_0 \subseteq \mathrm{PS}_1 \subseteq \dots \mathrm{PS}_n = S \: .
\end{equation}

\subsection{Two families of twists}
In eleven dimensions, the variety of square zero supercharges is described in
coordinates by the eleven equations
\begin{equation}
	\lambda^\alpha  \Gamma^\mu_{\alpha \beta} \lambda^\beta = 0 \: .
\end{equation}
This variety is closely related to the variety of Cartan pure spinors. In fact,
one finds $Y = \mathrm{PS}_3$~\cite{Eager:2018oww}. The variety of Cartan pure
spinors sits inside $Y$ as a subvariety $\mathrm{PS}_0 \subset \mathrm{PS}_3 =
Y$. Furthermore, $\mathrm{PS}_0$ is the singular locus of $Y$ and can be
described by imposing the additional equations
\begin{equation}
\lambda \Gamma^{\mu \nu} \lambda = 0 \: .
\end{equation}
For $Q$ on the singular locus, the degree zero part of the commutant is $Z^0(Q)
= \mathfrak{u}(5)$. This corresponds to the holomorphic twist of
eleven-dimensional supergravity. Away from the singular locus, the commutant is
an algebra with Levi factor $\mathfrak{g}_2 \times \mathfrak{gl}(2)$. This
corresponds to the maximal twist of eleven-dimensional supergravity that we will
study.

Let us elaborate a little further on the maximal twist. The spinor
representation in eleven dimensions decomposes as
\begin{equation}
	S_{11} = S_{7} \otimes S_{4} \: .
\end{equation}
The Dirac Spin representation in four dimensions, $S_{4},$ decomposes into Weyl
spinor representations $S_+$ and $S_-$:
\begin{equation}
	S_{4} = \wedge^\bullet L_{4}^\vee = \wedge^{\mathrm{even}} L_{4}^\vee \oplus \wedge^{\mathrm{odd}} L_{4}^\vee =: S_+ \oplus S_- \: .
\end{equation}
Identifying the group $Spin(4) \cong SU(2)_{+} \times SU(2)_{-},$ $S_{+}$ and
$S_{-}$ are the fundamental representations of $SU(2)_{+}$ and $SU(2)_{-},$
respectively. On a manifold $M^7$ with $G_2$ holonomy, the spinor representation
$S_{7}$ further decomposes as
\begin{equation}
	S_{7} = 1_{G_2} \oplus V_{G_2} \: ,
\end{equation}
where $V_{G_2}$ is the seven-dimensional representation of $G_2.$ Thus we have
the decomposition
\begin{equation}
	S_{11} = (1_{G_2} \oplus V_{G_2}) \otimes (\wedge^0 L_{4}^\vee \oplus \wedge^2 L_{4}^\vee \oplus S_-) \: .
\end{equation}
As a representation of $G_2 \times SU(2)_- \times U(1)_L$, where $U(1)_L$ is the
Cartan subgroup of $SU(2)_+$ this gives
\begin{equation}
	S_{11} = \left( (00) \oplus (10) \right) \otimes \left( 1_{-1} \oplus 1_{+1} \oplus 2_{0} \right) \: . 
\end{equation}
Here we introduced Dynkin labels for the $G_2$-representation. $SU(2)\times
U(1)$-representations are labeled by the dimension of the
$SU(2)$-representation, with the $U(1)$ charge as a subscript. To study the
maximal twist, we choose a square zero supercharge
\begin{equation}
Q \in 1_{G_2} \otimes \wedge^0 L_{4}^\vee = (00) 1_{-1} \: .
\end{equation}
Thus, we immediately see that $Q_{G_2}$ is invariant under the action of $G_2$
and $Q_{SU(2)_{+}}$ is invariant under the action of $SU(2)_{-}$ and has
$U(1)_L$ charge $-1,$.

The normal space to the nilpotence variety is spanned by the supercharges
\begin{align}
\Q_m & \in (V_{G_2} \otimes \wedge^2 L_{4}^\vee), \\
Q_{\dot{\alpha}} & \in (1_{G_2} \otimes S_{-}).
\end{align}
They satisfy the anticommutator relations
\begin{align}
\{Q , Q_m \} & = P_m \\
\{Q , Q_{\dot{\alpha}} \} & = P_{- \dot{\alpha}} \: .
\end{align}
Here we already used that the vector representation decomposes under $G_2 \times
SU(2) \times U(1)$ as
\begin{equation}
V_{11} = (10) \oplus 2_{-1} \oplus 2_1 \: .
\end{equation}
Our conventions are that indices $m,n, \dots$ are indices for the
seven-dimensional vector representation, while $\dot{\alpha},\dot{\beta},\dots$
correspond to $SU(2)_-$.

The above anticommutator relations state that translations on $M^7$ and
anti-holomorphic translations on $M^4$ are exact. Therefore, one can see already
at this point that the twisted theory will depend only on the topological
structures of $M^7$, but will be sensitive to holomorphic ones of $M^4$. Hence
this twist is ``partially topological'' or ``holomorphic-topological''.

\section{Eleven-dimensional supergravity in the pure spinor superfield formalism}
\label{sec:PS}
In this section, we give a short review of eleven-dimensional supergravity in
the pure spinor superfield formalism. The pure spinor superfield formalism has
been developed in the physics literature, in particular by
Berkovits~\cite{Berkovits:2001rb} and Cederwall~\cite{Cederwall:2013vba}. In the
context of eleven-dimensional supergravity, we in particular refer
to~\cite{Berkovits:2002uc, Cederwall:2010tn} and the references therein. The
pure spinor superfield formalism was reinterpreted from a more modern
mathematical perspective in~\cite{Eager:2018dsx}. For a detailed treatment in
modern language, we refer to our forthcoming work \cite{rfsw2021}. Here we only
use the pure spinor formalism as a tool to describe the action of supersymmetry
on the BV complex of eleven-dimensional supergravity and therefore only give a
brief treatment, exclusively tailored to the example of eleven-dimensional
supergravity.

\subsection{General remarks}
The general idea of the pure spinor superfield formalism is to replace the usual
BV complex $(\mathcal{E},D)$ by a much larger object, which we will denote by
$(A,\cD)$, encoding the same information. In the case of eleven-dimensional
supergravity, the construction can be carried out in the following way. Let $R=
\mathcal{O}(S_{11}) = \mathbb{C}[\lambda^1,\dots,\lambda^{32}]$ be the ring of
polynomial functions on the eleven-dimensional spin representation $S_{11}$ and
\begin{equation}
	I=(\lambda\Gamma^\mu \lambda)
\end{equation}
the ideal generated by the defining equation of the nilpotence variety. The
quotient $R/I$ can then be identified with the ring of functions on the
nilpotence variety $Y$. Furthermore, let $T$ denote the supertranslation
subgroup of the super Poincar\'e group. There are two commuting actions of $T$
on the smooth functions $C^\infty(T)$ on $T$, namely acting by left and right
translations. Infinitesimally, these actions are described by Lie algebra maps
\begin{equation}
	L,R : \mathfrak{t} \longrightarrow \mathrm{Vect}(T) \: .
\end{equation}
We denote the images of a basis of $\mathfrak{t}_-$ under $L$ and $R$ by
$\mathcal{Q}_\alpha$ and $\cD_\alpha$. Introducing coordinates $\theta^\alpha$
on $\mathfrak{t}_-$ and $x^\mu$ on $\mathfrak{t}_+= V_{11}$, these vector fields
are the usual left and right translations on superspace.
\begin{equation} \label{eq:QD}
\begin{split}
		\mathcal{Q}_\alpha &= \frac{\partial}{\partial \theta^\alpha} -  \Gamma^\mu_{\alpha \beta} \theta^\beta \frac{\partial}{\partial x^\mu}\\
		\cD_\alpha &= \frac{\partial}{\partial \theta^\alpha} + \Gamma^\mu_{\alpha \beta} \theta^\beta \frac{\partial}{\partial x^\mu} \: .
\end{split}
\end{equation}
Now we can define
\begin{equation} \label{total}
\left( A ,\cD \right) = \left( C^\infty(T) \otimes R/I \: , \: \mathcal{D} = \lambda^\alpha \mathcal{D}_\alpha \right) \: .
\end{equation}
Note that the defining equation of the ideal $I$ ensures that the differential
$\cD$ is indeed square zero. In coordinates, an element of this complex can be
thought of as a function $\Psi(x,\theta,\lambda)$ and is called a pure spinor
superfield. With these coordinates, $A$ takes the form
\begin{equation}
	\left( A ,\cD \right) =  \left( C^\infty(V_{11}) \otimes\C[\theta^1,\dots,\theta^{32}] \otimes R/I \: , \: \cD \right) \: .
\end{equation}
The differential $\cD$ has an obvious decomposition
\begin{equation}
	\cD = \cD_0 + \cD_1
\end{equation}
where $\mathcal{D}_0 = \lambda^\alpha \frac{\partial}{\partial \theta ^\alpha}$.
This makes $(A,\cD)$ into a bicomplex. Note that
\begin{equation}
	\left( A, \cD_0 \right) = \left( K^\bullet(R/I) \otimes C^\infty(V_{11}) \: , \: d_K \otimes \mathrm{id}_{C^\infty(V_{11})} \right) 
\end{equation}
is simply the Koszul complex of $R/I$ tensored with smooth functions on $V$.

The usual component field description is obtained by taking the cohomology with
respect to $\cD_0$ and transferring all relevant structures like the
differential $\cD_1$ (which gives to the differential $D$ on the component field
in the sense of Section~\ref{sec:SusyandBV}), the action of the supersymmetry
algebra (which gives rise to the $L_\infty$ module structure on the component
fields), and possibly higher terms of an $L_\infty$ structure (which would rise
to an $L_\infty$ structure encoding interactions on the component field level).
For a systematic account of this perspective, we refer the reader to our future
work~\cite{rfsw2021}. In the next subsection, we will explain how the
$\cD_0$-cohomology can be computed and how the action of supersymmetry can be
transferred for the case of eleven-dimensional supergravity.
\subsection{$\cD_0$-cohomology and representatives}
The $\cD_0$-cohomology can be identified with the tensor product
\begin{equation}
	(L^\bullet \otimes_R \C )  \otimes C^\infty(V_{11}) \: ,
\end{equation}
where $L^\bullet$ is the minimal free resolution of $R/I$ in $R$-modules. In our
case the minimal free resolution of $R/I$ takes the form
\begin{equation}
	\begin{tikzpicture}[descr/.style={fill=white,inner sep=1.5pt}]
	\matrix (m) [
	matrix of math nodes,
	row sep=2em,
	column sep=1.3em,
	text height=1.5ex, text depth=0.25ex
	]
	{R \otimes \Big( \C & V_{11} & \wedge^2 V_{11} \oplus V_{11} & \wedge^3 V_{11} \oplus \Sym^2(V_{11}) \oplus S_{11} & S_{11}\otimes V_{11} \\
		& S_{11} \otimes V_{11} & \wedge^3 V_{11} \oplus \Sym^2(V_{11}) \oplus S_{11} & \wedge^2 V_{11} \oplus V_{11} & \C \Big) \: .\\
	};
	
	\path[overlay,<-, font=\scriptsize,>=latex]
	(m-1-1) edge node[auto] {\(d_1\)} (m-1-2)
	(m-1-2) edge node[auto] {\(d_2\)} (m-1-3)
	(m-1-3) edge node[auto] {\(d_3\)} (m-1-4)
	(m-1-4) edge node[auto] {\(d_4\)} (m-1-5)
	(m-1-5) edge[out=355,in=175] node[yshift=1.3ex] {\(d_5\)} (m-2-2)
	(m-2-2) edge node[auto] {\(d_6\)} (m-2-3)
	(m-2-3) edge node[auto] {\(d_7\)} (m-2-4)
	(m-2-4) edge node[auto] {\(d_8\)} (m-2-5);
	\end{tikzpicture}
\end{equation}
The resolution differential was already described in~\cite{Berkovits:2002uc}.
Let us choose a basis $(e_\mu)$ of $V_{11}$ and $(s_\alpha)$ of $S_{11}$. We
will need the maps $d_1, \dots d_5$. In this basis they take the following form.
\begin{equation}
\label{eq:differentials}
\begin{matrix}
d_1&:& V_{11} \longrightarrow  \mathbb{C} & C^{(1)}& \mapsto& (\lambda \Gamma^\mu \lambda) C^{(1)}_\mu \\
&&&&& \\
d_2 &:& \wedge^2 V_{11} \oplus V_{11} \longrightarrow V_{11}& v &\mapsto& (\lambda \Gamma^{\mu \nu} \lambda) v_\mu e_\nu \\
&&&C^{(2)}&\mapsto& (\lambda\Gamma^\mu \lambda) C^{(2)}_{\mu \nu} e^\nu \\
&&&&& \\
d_3 &:& \wedge^3 V_{11} \oplus \mathrm{Sym}^2(V_{11}) \oplus S_{11} \longrightarrow \wedge^2 V_{11} \oplus V_{11}  &C^{(3)} &\mapsto& (\lambda \Gamma^\mu \lambda) C^{(3)}_{\mu \nu \rho} (e^\nu \wedge e^\rho) \\
&&&g&\mapsto& \left((\lambda \Gamma^\mu \lambda) e_\nu + \eta_{\rho \sigma}(\lambda\Gamma^{\sigma\nu}\lambda) (e^\mu \wedge e^\rho) \right) g_{\mu \nu}  \\
&&&\omega&\mapsto & \left( (\lambda \Gamma^\mu)_\alpha  e_\mu + \frac{1}{2}(\lambda\Gamma_{\mu \nu})_\alpha (e^\mu \wedge e^\nu )\right) \omega^\alpha\\
&&&&& \\
d_4 &:& S_{11}\otimes V_{11} \longrightarrow \wedge^3 V_{11} \oplus \mathrm{Sym}^2(V_{11}) \oplus S_{11} & \psi &\mapsto& -(\lambda \Gamma^\mu \lambda) \psi^\alpha_\mu s_\alpha +  \frac{1}{2}(\lambda \Gamma^{\mu \nu})_\alpha (\lambda \Gamma_\mu)^\beta \psi_{\nu \beta} s^{\alpha} \\
&&&&& + \frac{1}{2}(\lambda \Gamma_\mu)^\alpha \psi_{\nu \alpha} (e^{(\mu} \otimes e^{\nu)}) \\
&&&&&+ \frac{1}{4}(\lambda \Gamma_{\nu \rho})^\alpha \psi_{\mu \alpha} e^\mu \wedge e^\nu \wedge e^\rho \\
\\
d_5 &:& S_{11}\otimes V_{11} \longrightarrow S_{11} \otimes V_{11} & \psi^\dagger& \mapsto& (\lambda M_{\mu \nu}^{\alpha \beta} \lambda) \psi^{\dagger \nu}_\beta v^\mu \otimes s_\alpha.  \\
\end{matrix}
\end{equation}
We do not specify the tensor $M_{\mu \nu}^{\alpha \beta}$ here, but just remark
that it is a rather complicated  expression in terms of $\Gamma$-matrices. The
$\cD_0$-cohomology is bigraded by $\lambda$ and $\theta$. The component fields
organize according to degree in $\lambda$ and $\theta$ according to
Table~\ref{tab:BVmult}. We will call the $\lambda$ degree the BV degree.  This
convention is non-standard because it places the physical fields in BV degree
three. However, we will see that it simplifies other aspects of our
presentation.  
\begin{table}[htp]
\caption{$\theta$ and $\lambda$ degrees for the M-theory three-form BV multiplet}
\begin{center}
\begin{tabular}{|c||c|c|c|c|c|c|c|c|}
\hline
\diagbox{$\theta$}{$\lambda$} & 0 & 1 & 2 & 3 & 4 & 5 & 6 & 7 \\
 \hline
 \hline
0 &$C^{(0)}$ &   &  &  &  &  &  &  \\
 \hline
1 &  & $C^{(1)}$  &  &  &  &  &  &  \\
 \hline
2 & & & $C^{(2)}, v_{\mu}$    &  &  &  &  &  \\
\hline
3 &  &   &  $\omega$ & $C^{(3)}, g_{\mu \nu}$   &  &  &  &  \\
\hline
4 &  &   &   & $\psi$  &  &  &  &  \\
 \hline
5 &  &   &   & & $\psi^{\dagger}$   &  &  &   \\
 \hline
6 &  &   & &  &   $C^{(3)\dagger}, g^{\dagger}_{\mu \nu}$  &  $\omega^{\dagger}$
&  & \\
\hline
7 & & & &  &  &  $C^{(2)\dagger}, v_{\mu}^{\dagger}$  &  &  \\
\hline
8 &  & &  &  &  &  & $C^{(1)\dagger}$  &  \\
 \hline
9 &  &   &  &  &  &  &  & $C^{(0)\dagger}$  \\
 \hline
\end{tabular}
\end{center}
\label{tab:BVmult}
\end{table}
To find explicit representatives for the cohomology classes corresponding to the
component fields we define the adjoint differential
\begin{equation}
\mathcal{D}_0^\dagger = \theta^\alpha \frac{\partial}{\partial \lambda_\alpha} \: .
\end{equation}
Representatives can then be found by applying the resolution differential and
$\cD_0^\dagger$ iteratively. This was already noted in \cite{Movshev:2014hha}
and will be elaborated on in \cite{rfsw2021}. For example we find for the
one-form
\begin{equation}
C^{(1)} \xmapsto{d_1} (\lambda \Gamma^\mu \lambda) C^{(1)}_\mu \xmapsto{\mathcal{D}_0^\dagger} (\lambda \Gamma^\mu \theta) C^{(1)}_\mu \: ,
\end{equation}
such that the one-form field is represented by $(\lambda \Gamma^\mu \theta)
C^{(1)}_\mu$.

Similarly one finds for the two-form
\begin{equation}
C^{(2)} \xmapsto{d_2} (\lambda \Gamma^\mu \lambda) C^{(2)}_{\mu \nu} e^\nu \xmapsto{\mathcal{D}_0^\dagger} (\lambda \Gamma^\mu \theta) C^{(2)}_{\mu \nu} e^\nu \xmapsto{d_1} (\lambda \Gamma^\nu \lambda)(\lambda \Gamma^\mu \theta) C^{(2)}_{\mu \nu} \xmapsto{\mathcal{D}_0^\dagger} (\lambda \Gamma^\nu \theta)(\lambda \Gamma^\mu \theta) C^{(2)}_{\mu \nu} \: ,
\end{equation}
such that the two-form is represented by $(\lambda \Gamma^\nu \theta)(\lambda
\Gamma^\mu \theta) C^{(2)}_{\mu \nu}$. Likewise, the three-form is represented
by $(\lambda \Gamma^\nu \theta)(\lambda \Gamma^\mu \theta) (\lambda \Gamma^\rho
\theta) C^{(3)}_{\mu \nu \rho}$. \\ \\
Let us continue with the vector ghost $v$
\begin{equation}
v \xmapsto{d_2} (\lambda \Gamma^{\mu \nu} \lambda) v_\nu e_\mu \xmapsto{\mathcal{D}_0^\dagger} (\lambda \Gamma^{\mu \nu} \theta) v_\nu e_\mu \xmapsto{d_1} (\lambda \Gamma_\mu \lambda)(\lambda \Gamma^{\mu \nu} \theta) v_\nu \xmapsto{\mathcal{D}_0^\dagger} (\lambda \Gamma_\mu \theta)(\lambda \Gamma^{\mu \nu} \theta) v_\nu \: .
\end{equation}
Thus the representative is $(\lambda \Gamma_\mu \theta)(\lambda \Gamma^{\mu \nu}
\theta) v_\nu$. For the graviton we find with a similar calculation $(\lambda
\Gamma_\mu \theta)(\lambda \Gamma^{\mu(\nu} \theta)(\lambda
\Gamma^{\rho)}\theta) g_{\rho \nu}$. \\ \\
Performing this procedure one can find representatives for the gravitino and its
ghost. The results are summarized in Table~\ref{tab:11d reps}.
\begin{table}[h]
	\caption{Representatives for the fields in 11D supergravity organized by $\theta$-degree.}
	\begin{center}
		\begin{tabular}{|c|c|}
			\hline
			Field & Representative on the $E_1$-page \\
			\hline
			$C^{(0)}$  & $C^{(0)}$ \\
			\hline
			$C^{(1)}$ &$(\lambda \Gamma^\mu \theta) C^{(1)}_\mu$ \\
			\hline
			$C^{(2)}$ &$(\lambda \Gamma^\mu \theta)(\lambda \Gamma^\nu \theta)
			C^{(2)}_{\mu \nu}$ \\
			$v$ & $(\lambda \Gamma_\mu \theta)(\lambda \Gamma^{\mu \nu} \theta)
			v_\nu$ \\
			\hline
			$\omega$ & $\left[ (\lambda\Gamma_\mu\theta)(\lambda\Gamma^{\mu
			\nu}\theta)(\theta \Gamma_\nu)_\alpha + \frac{1}{2} (\lambda
			\Gamma^\mu \theta)(\lambda \Gamma^\nu \theta)(\theta \Gamma_{\mu
			\nu}) \right] \omega^\alpha$ \\
			$C^{(3)}$ & $(\lambda \Gamma^\mu \theta)(\lambda \Gamma^\nu
			\theta)(\lambda \Gamma^\rho \theta) C^{(3)}_{\mu \nu \rho}$ \\
			$g$ & $(\lambda \Gamma_\mu \theta)(\lambda \Gamma^{\mu(\nu}
			\theta)(\lambda \Gamma^{\rho)}\theta) g_{\rho \nu}$ \\
			\hline
			$\psi$ & $\left[ (\lambda\Gamma^\mu \theta)(\lambda\Gamma^\nu
			\theta)(\lambda\Gamma^\rho \theta)(\theta \Gamma_{\nu \rho})_\alpha
			- (\lambda \Gamma^\mu \theta)(\lambda \Gamma^{\nu\rho}
			\theta)(\lambda \Gamma_\nu \theta)(\theta \Gamma_\rho)_\alpha
			\right] \psi^\alpha_\mu$ \\
			\hline
		\end{tabular}
	\end{center}
	\label{tab:11d reps}
\end{table}
\subsection{The BV differential}
The differential $D$ acting on the component fields is obtained by transferring
$\cD_1$ to the $\cD_0$-cohomology. In general, this is done by a homotopy
transfer of $D_\infty$-algebras but here we are only interested in the lowest
order term that acts on the representatives simply by the usual formula of
$\cD_1$,
\begin{equation}
\mathcal{D}_1 = (\lambda \Gamma^\mu \theta) \partial_\mu \: .
\end{equation}
This gives part of the differential, that is first order in derivatives. For
example, we can act on the $C^{(0)}$ ghost
\begin{equation}
\mathcal{D}_1 (C^{(0)}) = (\lambda \Gamma^\mu \theta) \partial_\mu C^{(0)} \: .
\end{equation} 
Thus we see that the differential corresponds to the de Rham differential. This
obviously generalizes to $C^{(1)}$ and $C^{(2)}$ such that we see that the ghost
system of the three-form indeed corresponds to the usual ghost system of a
higher form field. Moving on to the diffeomorphism ghost $v_\mu$ for the
graviton, we find
\begin{equation}
\mathcal{D}_1( (\lambda\Gamma_\mu \theta) (\theta \Gamma^{\mu \nu} \theta)v_\nu) = (\lambda\Gamma_\mu \theta) (\theta \Gamma^{\mu \nu} \theta) (\lambda \Gamma^\rho \theta) \partial_\rho v_\nu \: .
\end{equation}
From our calculations of the representatives, we know that only the part where
$\rho$ and $\nu$ are symmetrized corresponds to a non-trivial cohomology class.
Thus we find
\begin{equation}
\mathcal{D}_1(v) = (\lambda\Gamma_\mu \theta) (\theta \Gamma^{\mu (\nu} \theta) (\lambda \Gamma^{\rho)} \theta) (\partial_\rho v_\nu + \partial_\nu v_\rho) \: .
\end{equation}
Written dually in terms of operators, we find that the BV operator acts by
\begin{equation}
Q_{BV} g_{\mu \nu} = \partial_\mu v_\nu + \partial_\nu v_\mu = (\mathcal{L}_v \eta)_{\mu \nu} \: ,
\end{equation}
which is indeed the expected gauge transformation for the graviton.

A similar story also holds for the gravitino and its ghost. There we find
\begin{equation}
	\cD_1(\omega) = (\lambda \Gamma^\rho \theta)\left[ (\lambda\Gamma_\mu\theta)(\lambda\Gamma^{\mu \nu}\theta)(\theta \Gamma_\nu)_\alpha + (\lambda \Gamma^\mu \theta)(\lambda \Gamma^\nu \theta)(\theta \Gamma_{\mu \nu})_\alpha \right] \partial_\rho \omega^\alpha \: .
\end{equation}
This gives a gauge transformation
\begin{equation}
	Q_{BV} \psi_\mu^\alpha = \partial_\mu \omega^\alpha \: .
\end{equation}
Thus we see that $\cD_1$ encodes the usual gauge transformations, expected for
the field content. Furthermore, one expects $\cD_1$ to encode the
Rarita--Schwinger equation between the gravitino and its antifield. In addition,
homotopy transfer is expected to induce a second order differential giving the
linearized equations of motions of the graviton and the three-form field.

\subsection{The action of supersymmetry} 
\label{sec:SusyTrafos}
As explained in Section~\ref{sec:SusyandBV}, the supersymmetry algebra usually
does not act strictly on the component fields. This is indeed the case for
eleven-dimensional supergravity. Instead there is a $L_\infty$ map
\begin{equation}
\rho: \mathfrak{p} \rightsquigarrow \cD(\mathcal{E}) \: .
\end{equation}
The components of $\rho$ can be obtained from the action of $Q_\alpha$ by left
translation using a homotopy transfer procedure. The strict part is simply
obtained by letting~(\ref{eq:QD}) act on the representatives. For an element $Q=
\epsilon^\alpha Q_\alpha$ of the supersymmetry algebra, this means
\begin{equation}
\rho^{(1)}(Q) = \epsilon^\alpha\frac{\partial}{\partial \theta^\alpha} - \epsilon^\alpha \Gamma^\mu_{\alpha \beta} \theta^\beta \partial_\mu \: .
\end{equation}
For the second order part one finds
\begin{equation}
\rho^{(2)}(Q_1,Q_2) = p \circ \left(\rho^{(1)}(Q_1) \circ \mathcal{D}_0^\dagger \circ \rho(Q_2) + \rho(Q_2) \circ \mathcal{D}_0^\dagger \circ \rho^{(1)}(Q_1) \right) \circ i \: .
\end{equation}
Here $i$ is the inclusion map from the $\cD_0$-cohomology to the total
complex~(\ref{total}) (mapping a component field to its representative, as
computed above) and $p$ is the projection back onto the $\cD_0$-cohomology. The
presence of $\rho^{(2)}$ signals that the supersymmetry transformations only
close up to the equations of motions and gauge transformations. In fact,
$\rho^{(2)}$ nullhomotopes the failure of the supersymmetry algebra to be
represented strictly and thus exactly corresponds to what is called a ``closure
term'' in the physics literature. Higher order components will not appear for
eleven-dimensional supergravity. In~\cite{Berkovits:2002uc}, a close connection
between the resolution differential and the non-derivative supersymmetry
transformations and their closure terms was conjectured. We will explain this
claim in modern language and provide a proof in~\cite{rfsw2021}. Here it
suffices to say that, as we will see momentarily, one can obtain the
non-derivative part of $\rho$ by replacing $\lambda$ with $\epsilon$ in the
resolution differential at appropriate places. Now let us start deriving the
action of the supersymmetry algebra on the BV fields. The strict part gives the
usual supersymmetry transformations known from the
literature~\cite{Berkovits:2002uc}.
\subparagraph{The three-form ghost system}
We begin with the ghost system of the three-form. From degree reasons, it is
obvious that $\rho^{(1)}$ acts trivially on the ghost system for the three-form.
Thus we have
\begin{equation}
\rho^{(1)}(C^{(0)}) = \rho^{(1)}(C^{(1)}) = \rho^{(1)}(C^{(2)}) = 0 \: .
\end{equation}
However, this will be corrected by higher order contributions. There we find
\begin{equation}
\begin{split}
\rho^{(2)}(Q,Q)(C^{(1)}) =& \rho^{(2)}(Q,Q)(C^{(1)}_\mu (\lambda \Gamma^\mu \theta)) \\
=&(\epsilon \Gamma^\mu \epsilon) C^{(1)}_\mu \\
=& \iota_{\{Q,Q\}} C^{(1)} \: .
\end{split}
\end{equation}
Thus we find a map
\begin{equation}
\rho^{(2)}(Q,Q) = \iota_{\{Q,Q\}} : \Omega^1(M) \longrightarrow \Omega^0(M) \: .
\end{equation}
Here we also see the relation to the resolution differential: $d_1$ acts on the
one-form by $C^{(1)} \rightarrow (\lambda \Gamma^\mu \lambda) C^{(1)}_\mu,$ thus
replacing $\lambda$ with $\epsilon$ we obtain $\rho^{(2)}$.

Written dually for operators, this gives a supersymmetry transformation rule
\begin{equation}
	\delta C^{(1)}_\mu = (\epsilon \Gamma_\mu \epsilon) C^{(0)} \: .
\end{equation}

With similar calculations, we also see that there are higher order
transformations
\begin{equation}
\rho^{(2)}(Q,Q) = \iota_{\{Q,Q\}} : \Omega^2(M) \longrightarrow \Omega^1(M)
\end{equation}
and
\begin{equation}
\rho^{(2)}(Q,Q) = \iota_{\{Q,Q\}} : \Omega^3(M) \longrightarrow \Omega^2(M) \: .
\end{equation}
However, these transformations will not cancel any components in the twist since
there the relevant supercharge satisfies $\{Q,Q\} = 0$ and thus the above maps
all vanish.
\subparagraph{The diffeomorphism ghost}
The only non-derivative transformation for the diffeomorphism ghost appears in
$\rho^{(2)}$. It takes the form
\begin{equation}
\begin{split}
\rho^{(2)}(Q,Q)(v) &= \rho^{(2)}(Q,Q)((\lambda \Gamma_\mu \theta)(\theta \Gamma^{\mu \nu} \lambda)v_\nu) \\
&= (\lambda \Gamma_\mu \theta) (\epsilon \Gamma^{\mu \nu} \epsilon) v_\nu
\end{split}
\end{equation}
and thus gives a transformation rule
\begin{equation}
	\delta C^{(1)}_\mu = (\epsilon \Gamma_{\mu \nu} \epsilon) v^\nu.
\end{equation}
In addition, there is a $\rho^{(1)}$-piece involving a derivative that can be
seen to give rise to the usual supersymmetry transformation between the
diffeomorphism and supertranslation ghost \cite{Berkovits:2002uc}
\begin{equation}
	\delta \omega_\alpha = - \frac{1}{2} (\epsilon
	\Gamma^{\mu \nu})_\alpha  \partial_\mu v_\nu \:.
\end{equation}

\subparagraph{The gravitino ghost}
For the gravitino ghost, we obtain
\begin{equation}
	\rho^{(1)}(Q)(\omega) = (\lambda \Gamma_\mu \theta)(\lambda \Gamma^{\mu \nu} \theta) (\epsilon \Gamma_\nu \omega) + \frac{1}{2} (\lambda \Gamma_\mu \theta) (\lambda \Gamma_\nu \theta) (\epsilon \Gamma^{\mu \nu} \omega) \: .
\end{equation}
Again, note the relation to the free resolution. This gives two supersymmetry
transformations
\begin{equation}
\begin{split}
	\delta v_\mu &= \epsilon \Gamma_\mu \omega \\
	\delta C^{(2)}_{\mu \nu} &= \frac{1}{2} \epsilon \Gamma_{\mu \nu} \omega \: .
\end{split}
\end{equation}

By now the methodology should be clear. In this style, one can derive the full
higher order corrections to the supersymmetry transformations and encode them in
the differential $\delta$.

We summarize the full non-derivative supersymmetry transformations in
Table~\ref{tab:11d trafos}. These results first appeared in
\cite{Berkovits:2002uc}.
\begin{table}[h]
	\caption{Non-derivative supersymmetry transformations}
	\begin{center}
		\begin{tabular}{|c|c|}
			\hline
			Operator $\phi$ & Transformation rule $\delta \phi$ \\
			\hline
			$C^{(0)}$  & $\delta C^{(0)} = (\epsilon \Gamma^\mu \epsilon)
			C^{(1)}_\mu$ \\
			\hline
			$C^{(1)}$ &$\delta C^{(1)}_\mu = (\epsilon \Gamma^\nu
			\epsilon)C^{(2)}_{\mu \nu} + (\epsilon \Gamma_{\mu \nu} \epsilon)
			v^\nu$ \\
			\hline
			$C^{(2)}$ &$\delta C^{(2)}_{\mu \nu} = \frac{1}{2} \epsilon
			\Gamma_{\mu \nu} \omega + (\epsilon \Gamma^\rho
			\epsilon)C^{(3)}_{\mu \nu \rho} + (\epsilon \Gamma_{[\mu \rho}
			\epsilon) g^{\rho}_{\ \nu]}$ \\
			$v$ & $\delta v_\mu = \epsilon\Gamma_\mu \omega + (\epsilon
			\Gamma^\nu \epsilon) g_{\mu \nu}$ \\
			\hline
			$\omega$ & $ \delta \omega_\alpha =  (\epsilon \Gamma^\mu \epsilon)
			\psi_{\alpha \mu} + \frac{1}{2} (\epsilon \Gamma^{\mu \nu})_{\alpha}
			(\epsilon \Gamma_{\mu})^{\beta} \psi_{\beta \nu}$ \\
			$C^{(3)}$ & $\delta C^{(3)}_{\mu \nu \rho} = \frac{1}{4} \epsilon
			\Gamma_{[\mu \nu} \psi_{\rho]}$ \\
			$g$ & $\delta g_{\mu \nu} = \frac{1}{2} \epsilon \Gamma_{(\mu}
			\psi_{\nu )}$ \\
			\hline
			$\psi$ & $\delta \psi^\alpha_\mu = (\epsilon M_{\mu \nu}^{\alpha
			\beta} \epsilon) \psi^{\dagger \nu}_{\beta}$ \\
			\hline
		\end{tabular}
	\end{center}
	\label{tab:11d trafos}
\end{table}
In addition, we list the transformations including derivatives for the gravitino
and its ghost in Table~\ref{tab:11d trafos deriv}.
\begin{table}[h]
	\caption{Supersymmetry transformations with derivatives}
	\begin{center}
		\begin{tabular}{|c|c|}
			\hline
			Operator $\phi$ & Transformation rule $\delta \phi$ \\
			\hline
			$\omega$ & $ \delta \omega_\alpha = (\epsilon \Gamma^{\mu
			\nu})_\alpha \partial_\mu v_\nu$ \\
			\hline
			$\psi$ & $\delta \psi^\alpha_\mu = (\Gamma^{\nu \rho \sigma
\tau}_{\mu} - 8 \Gamma^{\rho \sigma \tau} \delta_{\mu}^{\nu} )G^{(4)}_{\nu \rho
\sigma \tau} \epsilon^{\alpha} $ \\
			\hline
		\end{tabular}
	\end{center}
	\label{tab:11d trafos deriv}
\end{table}
\section{Twisting the free theory}
\label{sec:twistedfields}

In this section, we will show that the fields of the twisted theory arrange into
a differential form
\begin{equation}
	\cA \in \Omega^{\bullet}(M^7) \otimes \Omega^{0,\bullet}(M^4) \: .
\end{equation}
The strategy to establish this result is clear: we restrict the supersymmetry
transformations from Table~\ref{tab:11d trafos} to our $G_2 \times SU(2)$
invariant supercharge and look for fields that form trivial pairs under
$\delta$. In the twisted theory these fields decouple and can be neglected. To
find such cancellations we have to decompose the field content as well as the
supersymmetry transformations equivariantly under $G_2 \times SU(2) \times
U(1)$.

As a result, we will see that only certain components of the three-form, the
three-form ghost system, the gravitino, and the corresponding antifields play a
role in the twisted theory. These fields then arrange into the differential form
described above. 

We will see that the twisted differential takes the form
\begin{equation}
\label{eq:BVdifferential}
D^{\mathrm{tw}} = d_{M^7} \otimes 1 + 1 \otimes \delbar_{M^4}.
\end{equation}
The fields in the untwisted theory have a have a $\Z \times \Z$-grading given by
the BV degree $d_{BV}$ and the $U(1)_L$ charge $d_{U(1)_L}.$ After twisting, the
new BV operator $Q_{BV} + \delta_Q$ breaks the $\Z \times \Z$-grading on the
space of fields $E$ to the $\Z$-grading
\begin{equation}
\label{eq:new degree}
d^Q_{BV} = d_{BV} - d_{U(1)_L} ,
\end{equation}
in the twisted theory. Note that $D^{\mathrm{tw}}$ is not homogenous with
respect to this grading since $\delbar_{M^4}$ operator carries $U(1)_L$ charge
-1. The new BV degree of a component of $\cA$ is simply its de Rham form degree
on $M^7$. Alternatively, note that the twisted BV differential preserves the
total form degree and we can assign a total form degree to the components of
$\cA.$ We observe that for component fields in $\cA$ the total form degree
agrees with their original $\theta$-degree. However, interactions might not
preserve these degrees.
\subsection{Decomposition of the field content}

We now decompose the field content into representations of $G_2 \times SU(2)_-
\times U(1)_L$. To do this, recall the following sequence of inclusions
\begin{equation}
	Spin(11) \supset Spin(7) \times SU(2)_- \times U(1)_L \supset G_2 \times SU(2)_- \times U(1)_L \: .
\end{equation}
The branching of the relevant representations from $Spin(11)$ to $Spin(7) \times
SU(2)_- \times U(1)_L$ is described by Table~\ref{tab:SObranching}.
\begin{table}[htp]
	\caption{Branching of $Spin(11) \rightarrow Spin(7) \times SU(2)_{-} \times
	U(1)_L$-representations.}
	\begin{center}
		\begin{tabular}{|c|c|}
			\hline
			$Spin(11)$ & $Spin(7) \times SU(2)_- \times U(1)_L$ \\
			\hline
			(00000) & $(000)(1_0)$ \\ 
			\hline
			(10000)& $(000)(2_{-1} + 2_1) \oplus (100) (1_0) $ \\
			\hline (00001)& $(001) (1_{-1} + 1_1 + 2_0)$ \\
			\hline (01000) & $(000)(1_{-2} + 1_0 + 3_0 + 1_2) \oplus (010)(1_0)
			\oplus (100)(2_{-1} + 2_1)$ \\
			\hline (00100) & $(000)(2_{-1} + 2_1) \oplus (002)(1_0) \oplus
			(010)(2_{-1} + 2_1) \oplus (100)(1_{-2} + 1_0 + 3_0 + 1_2)$ \\
			\hline (20000) & $(000)(3_{-2} + 1_0 + 3_0 + 3_2) \oplus
			(100)(2_{-1} + 2_1) + (200)(1_0)$ \\
			\hline (10001) & $(001)(2_{-2} + 3_{-1} + 1_{-1} + 2_0^{\oplus 2}+
			1_1 + 3_1 + 2_2) \oplus (101)(1_{-1} + 2_0 + 1_1) $\\
			\hline
		\end{tabular}
	\end{center}
	\label{tab:SObranching}
\end{table}
Here we are using Dynkin labels to identify the $Spin(11)$ and $Spin(7)$
representations. We identify $SU(2) \times U(1)$-representations by the
dimension of the $SU(2)$-representation and denote the $U(1)$ charge as a
subscript. Recall that the vector representation $V_{11}$ has Dynkin label
$(10000)$ and its second and third exterior powers are labeled by $(01000)$ and
$(00100)$. The spinor representation $S_{11}$ has Dynkin label $(00001)$.
Furthermore, the gravitino representation already decomposes as a $Spin(11)$
representation according to
\begin{equation}
	S_{11} \otimes V_{11} \cong (00001) \oplus (10001) \: .
\end{equation}
Finally, the graviton transforms in the representation
\begin{equation}
	\Sym^2 V_{11} \cong (20000) \oplus (00000) \: .
\end{equation}
We also need the branching rules for $Spin(7) \rightarrow G_2,$ which we collect
in Table~\ref{tab:SObranching2}.
\begin{table}[htp]
	\caption{Branching of $Spin(7) \rightarrow G_2$-representations.}
	\begin{center}
		\begin{tabular}{|c|c|}
			\hline
			$Spin(7)$ & $G_2$ \\
			\hline
			(000) & $(00)$ \\ 
			\hline
			(100)& $(10)$ \\
			\hline 
			(001)& $(10) \oplus (00)$ \\
			\hline 
			(010) & $(01) \oplus (10)$ \\
			\hline 
			(002) & $(00) \oplus (10) \oplus (20)$ \\
			\hline 
			(101) & $(01) \oplus (10) \oplus (20)$ \\
			\hline
			(200) & $(20)$ \\
			\hline
		\end{tabular}
	\end{center}
	\label{tab:SObranching2}
\end{table}
We see that the three-form and its ghosts $C^{(p)}$ split into forms in
$\Omega^i(M^7) \otimes \Omega^{j_1,j_2}(M^4),$ where $i + j_1 +j_2 = p$ is the
total form degree. Thus, in the light of the conjecture, we expect all
components with non-zero holomorphic form degree ($j_1 \neq 0$) to cancel in the
twisted theory.

We now consider the decomposition of the gravitino field $\psi_{\mu}^{\alpha}$.
It transforms in the product of the $Spin(11)$  vector and spinor
representations. We first consider its decomposition under $Spin(11) \rightarrow
Spin(7) \times SU(2)_{-}$. We will later see that the only components that
survive in the twisted multiplet have index $\mu$ transforming in a
$Spin(7)$-vector representation whose components we denote by $m.$

On a manifold of $G_2$ holonomy the de Rham complex splits into three
sub-complexes~\cite{MR2292510}
\begin{equation}
\begin{tikzcd}
\Omega^0_1 \arrow[r,"d"] & \Omega^1_7 \arrow[r,"d"] & \Omega^2_7 \arrow[r,"d"] & \Omega^3_1 & \Omega^4_1 \arrow[r,"d"] & \Omega^5_7 \arrow[r,"d"] & \Omega^6_7 \arrow[r,"d"] & \Omega^7_1 \\
& & \Omega^2_{14} \arrow[r,"d"] \arrow[dr,"d"] & \Omega^3_7 \arrow[r,"d"]  & \Omega^4_7 \arrow[r,"d"] & \Omega^5_{14}.  & &  \\
& &  & \Omega^3_{27} \arrow[r,"d"]  & \Omega^4_{27} \arrow[ur,"d"] &   & &  \\
\end{tikzcd}
\end{equation}
To define the space of differential forms $\Omega^k_l$, recall that we can
identify the differential forms $\Omega^k$ in the de Rham complex with sections
of the $k$-th exterior power of the cotangent bundle of $M^7.$ When $M^7$ has
$G_2$ holonomy the exterior powers of the cotangent bundle decompose into
irreducible representations of $G_2$ and we denote the sections by $\Omega^k_l$,
where the subscript denotes the respective dimension of the
$G_2$-representation.

The spin 1/2 and spin 3/2 fields on $M^7$ decompose as \cite{Guio:2017zfn,
Homma:2018hmn}
\begin{align}
\Sigma_{1/2} & \cong \Omega^0_1 \oplus \Omega^1_7 \\
\Sigma_{3/2} & \cong \Omega^1_7 \oplus \Omega^2_{14} \oplus \Omega^3_{27}. 
\end{align}
Using the above decomposition and the $Spin(11) \rightarrow Spin(7) \times SU(2)
\times U(1)_L$ branchings in Table~\ref{tab:SObranching}, and the isomorphisms
\begin{align}
\Sigma_{3/2} \oplus \Sigma_{1/2} & \cong \left(\Omega^1_7 \oplus \Omega^2_{14} \oplus \Omega^3_{27} \right) \oplus \left( \Omega^0_1 \oplus \Omega^1_7  \right) \\
& \cong \Omega^2 \oplus \Omega^3,
\end{align}
we see that the gravitino, given by a pair of spin 3/2 and spin 1/2 fields on a
$G_2$ holonomy manifold, can be identified with a pair of two- and three-forms
on the manifold. We will find that the components of the gravitino that survive
the twist are contained in the representation
\begin{equation}
	(S_{+} \oplus S_{-}) \otimes (\Sigma_{3/2} \oplus \Sigma_{1/2} ) \cong (S_{+} \oplus S_{-}) \otimes (\Omega^2 \oplus \Omega^3) \: .
\end{equation} 
However, not all of these components survive. We will find that the surviving
components are $\Omega^3 \otimes \wedge^0 L_{4}^\vee, \Omega^3 \otimes S_{-}$,
and $\Omega^2 \otimes \wedge^2 L_{4}^\vee.$ The gravitino has BV degree 3 in the
untwisted theory and the representations $\wedge^0 L_{4}^\vee, S_{-}, \wedge^2
L_{4}^\vee$ have $U(1)$ charge $-1,0,$ and $1,$ respectively. Thus their new BV
defined by Equation~\eqref{eq:new degree} are 4, 3, and 2. The components
surviving the twist are therefore in $\Omega^4(M^7) \otimes \Omega^{0,0}(M^4)$,
$\Omega^3(M^7) \otimes \Omega^{0,1}(M^4),$ and $\Omega^2(M^7) \otimes
\Omega^{0,2}(M^4),$ where we have used the isomorphism $ \Omega^3 \cong
\Omega^4$ to ensure that the gravitino has its correct twisted BV degree. 
  
The components of the three-form and its ghosts $C^{(p)}$, $p = 0 \dots 3$ and
the gravitino along with their antifields that survive the twist therefore give
exactly the right field content to be described by a form
\begin{equation}
\cA \in \Omega^{\bullet}(M^7) \otimes \Omega^{0,\bullet}(M^4).
\end{equation}

\subsection{Decomposition of the supersymmetry transformations}
We now determine the supersymmetry transformations for the scalar supercharge
$Q$. For the moment we are only interested in the supersymmetry transformations
without derivatives since these are the ones responsible for the formation of
trivial pairs.  The transformations with derivatives will later be used to
determine the twisted BV differential. Recall that the spin representation
$S_{11}$ decomposes as
\begin{equation}
	\left[ (00) \oplus (10) \right] (1_{-1} + 1_1 + 2_0) \: .
\end{equation}
This means that we can decompose the parameter $\epsilon$ from
Table~\ref{tab:11d trafos} into
\begin{equation} \label{eq:eps decomp}
	\epsilon \rightarrow (\epsilon_{-} , \epsilon_{+}, \epsilon_{\dot{\alpha}} , \epsilon_{-m} , \epsilon_{+m}, \epsilon_{m \dot{\alpha}}) \: .
\end{equation}
Here $m$ is an index for the seven-dimensional representation of $G_2$. To act
by $Q,$ we specify $\epsilon_{-} = 1$ and set all other components to zero. 

On general grounds, these transformation take a very simple form.  As explained
above, the supercharge $Q$ is invariant under $G_2 \times SU(2)$ and has $U(1)$
charge $-1$. As a consequence, $\delta_Q$ is an $G_2 \times SU(2)$-equivariant
map. By decomposing the field content into irreducible $G_2 \times
SU(2)$-representations, $\delta_Q$ splits up as a map between these
irreducibles. However, since $\delta_Q$ is equivariant, we can apply Schur's
lemma and find, first, that there can not be any non-trivial maps between
non-isomorphic components and, second, transformations between isomorphic $G_2
\times SU(2)$-representations are always of the form $\alpha \cdot \mathrm{id}$
for some $\alpha \in \C$. Thus, to check whether there are any trivial pairs, we
only have to see if there is a non-vanishing map between isomorphic
representations. In addition, $\delta_Q$ carries a $U(1)$ charge that simply
equals minus the number of $\epsilon$'s appearing in the transformation, which
can be used as a further criterion to establish that certain maps vanish.

To check whether or not supersymmetry transformation yields a trivial pair we
need to decompose $\Gamma$-matrices.

\subparagraph{Gamma matrix decomposition}
In eleven dimensions the symmetric square of the spin representation decomposes
as
\begin{equation}
\Sym^2 S_{11} \cong V_{11} \oplus  \wedge^2 V_{11} \oplus  \wedge^5 V_{11} \: .
\end{equation}
Accordingly, there are maps denoted by $\Gamma^\mu$, $\Gamma^{\mu \nu}$ and
$\Gamma^{\mu_1 \dots \mu_5}$ given by projecting onto the summands in this
decomposition. So for example, $\Gamma^\mu$ is given by the composition
\begin{equation}
\begin{tikzcd}
\Sym^2(S_{11}) \arrow[r,"\cong"] \arrow[dr,"\Gamma^\mu"] & V_{11} \oplus  \wedge^2 V_{11} \oplus  \wedge^5 V_{11} \arrow[d] \\
& V_{11} \\
\end{tikzcd} \: .
\end{equation}
Recall the spin representation $S_{11}$ decomposes under $G_2 \times SU(2)
\times U(1)$ as
\begin{equation}
S_{11} \rightarrow1_{-1} + 1_{1} + 2_0 + (10) (1_{-1} + 1_1 + 2_0) \: .
\end{equation}
We are interested in $\epsilon_- \Gamma^\mu \epsilon$ and $\epsilon_-
\Gamma^{\mu\nu} \epsilon$, where $\epsilon_- \in 1_{-1}$ in the above
decomposition and $\epsilon$ is arbitrary. This means we are looking at a map
$1_{-1} \otimes S_{11} \rightarrow V_{11}$ or $1_{-1} \otimes S_{11} \rightarrow
\wedge^2 V_{11},$ respectively. The representations $V_{11}$ and $\wedge^2
V_{11}$ decompose as
\begin{equation}
\begin{split}
V_{11} &\rightarrow 2_1 \oplus 2_{-1} \oplus (10) \\
\wedge^2 V_{11} &\rightarrow (1_{-2} \oplus 1_0 \oplus 3_0 \oplus 1_2) \oplus (10)(2_{-1} \oplus 2_1) \oplus (10) \oplus (01). 
\end{split} 
\end{equation}
We can now compare this with the decomposition of $1_{-1} \otimes S_{11}$ and
read off the following results for $\Gamma^\mu$:
\begin{equation} \label{eq:gamma decomp}
\begin{matrix}
\epsilon_- \Gamma^\mu \epsilon_- & = & 0 \\
\epsilon_- \Gamma^\mu \epsilon_+ & = & 0 \\
\epsilon_- \Gamma^\mu \epsilon_{\dot{\alpha}} & \in & 2_{-1} \\
\epsilon_- \Gamma^\mu \epsilon_{+ m} & \in & (10)  \\
\epsilon_- \Gamma^\mu \epsilon_{-m} & = & 0 & \\
\epsilon_- \Gamma^\mu \epsilon_{m \dot{\alpha}} & = & 0. & \\
\end{matrix}
\end{equation}
For $\Gamma^{\mu \nu}$ we find:
\begin{equation} \label{eq:gamma2 decomp}
\begin{matrix}
\epsilon_- \Gamma^{\mu \nu} \epsilon_- & \in & 1_{-2}  \\
\epsilon_- \Gamma^{\mu \nu} \epsilon_+ & \in & 1_0 \\
\epsilon_- \Gamma^{\mu \nu} \epsilon_{\dot{\alpha}} & = & 0 \\
\epsilon_- \Gamma^{\mu \nu} \epsilon_{+ m} & = & 0  \\
\epsilon_- \Gamma^{\mu \nu} \epsilon_{-m} & = & 0 & \\
\epsilon_- \Gamma^{\mu \nu} \epsilon_{m \dot{\alpha}} & \in & (10) \: 2_{-1}. \\
\end{matrix}
\end{equation}
For example, we immediately see that all terms of the form $\epsilon_-
\Gamma^\mu \epsilon_-$ vanish and hence do not affect the twist. This is also a
direct consequence of $Q$ being nilpotent.

Let us start examining the supersymmetry transformations. Note that we are
ignoring any potential non-zero scalar coefficients $\alpha$ as we are only
interested in the formation of trivial pairs.

Furthermore, we are only considering cancellations between the fields of the
multiplet as well as between the gravitino and its antifield. Since the action
of supersymmetry respects the pairing on the BV complex, the same cancellations
also occur for the respective antifields. 
\subparagraph{The zero-form $C^{(0)}$}
For the zero-form ghost, we obviously have $\delta_Q C^{(0)} = 0$. Since there
is no supersymmetry transformation generating $C^{(0)}$, it will become a field
in the twisted theory.
\subparagraph{The diffeomorphism ghost $v$}
Next we consider the diffeomorphism ghost $v_\mu$. It decomposes into components
\begin{equation}
 v_\mu \rightarrow (v_{m} , v_{+ \dot{\alpha}}, v_{- \dot{\alpha}}) \: .
\end{equation}
We have a supersymmetry transformation of the form
\begin{equation}
	\delta_Q v_\mu = \epsilon \Gamma_\mu \omega \: .
\end{equation}
The gravitino ghost $\omega$ lives in the spinor representation and hence
decomposes according to Equation~\eqref{eq:eps decomp}. From the $\Gamma$-matrix
decomposition in Equation~\eqref{eq:gamma decomp}, we know that $\epsilon_-
\Gamma_\mu \omega$ is only non-vanishing for the components
$\omega_{\dot{\alpha}}$ and $\omega_{+m}$ of $\omega$. Thus we get up to
potential non-zero prefactors
\begin{equation}
	\delta_Q v_{m} = \omega_{+m}
\end{equation}
and
\begin{equation}
	\delta_Q v_{-\dot{\alpha}} = \omega_{\dot{\alpha}} \: .
\end{equation}
 Finally we have,
\begin{equation}
\delta_Q v_{+ \dot{\alpha}} = 0 \: .
\end{equation}
Thus we already find that some components of the diffeomorphism ghost $v$ form
trivial pairs with parts of the gravitino ghost. In addition, it is interesting
to note that $\delta_Q v_{+ \dot{\alpha}} = 0$. As $v_{+ \dot{\alpha}}$ will not
be part of the twisted three-form multiplet, we expect it to be in the image of
$\delta_Q$, forming a trivial pair with another field.  Indeed, we will
momentarily find that $v_{+ \dot{\alpha}}$ cancels the holomorphic part of the
one-form $C^{(1)}$. 
\subparagraph{The one-form $C^{(1)}$}
For the field $C^{(1)},$ we have a supersymmetry transformation rule
\begin{equation}
	\delta_Q C^{(1)}_\mu = (\epsilon_- \Gamma_{\mu \nu} \epsilon_-) v^\nu \: .
\end{equation}
From the $\Gamma$-matrix decomposition, we know $\epsilon_- \Gamma_{\mu \nu}
\epsilon_- \in 1_{-2}$. Thus we immediately find
\begin{equation}
	\delta_Q C^{(1)}_{m} = 0
\end{equation}
and
\begin{equation}
\delta_Q C^{(1)}_{+ \dot{\alpha}} = 0 \: .
\end{equation}
In addition, we have
\begin{equation}
\delta_Q C^{(1)}_{- \dot{\alpha}} = v_{+ \dot{\alpha}} \: .
\end{equation}
This shows that $C^{(1)}_{- \dot{\alpha}}$ and $v_{+ \dot{\alpha}}$ form a
trivial pair and thus do not appear in the twisted theory. Recall that the
choice $(\epsilon_{-} , \epsilon_{+} , \epsilon_{\dot{\alpha}}) = (1,0,0)$
defines a complex structure on $\R^4 \cong \C^2$. The four-dimensional vector
representation decomposes as
\begin{equation}
	V_4 = S_+ \otimes S_- = 2_1 \oplus 2_{-1} \: .
\end{equation}
The representation $2_{-1}$ corresponds to holomorphic and $2_1$ to the
antiholomorphic components. Thus we see that, for this complex structure, the
components $C^{(1)}_{- \dot{\alpha}}$ form the holomorphic parts of the one-form
ghost $C^{(1)}$. As expected, only the anti-holomorphic part of the one-form
plays a role in the twisted theory.

We can alternatively describe the cancellation using holomorphic geometry. With
respect to the complex structure on $\C^2$,
\begin{equation}
	\Omega = (\epsilon_- \Gamma_{\mu \nu} \epsilon_-) dx^\mu \wedge dx^\nu
\end{equation}
defines a holomorphic $(2,0)$-form. Introducing coordinates  $(z^{\dot{\alpha}},
\bar{z}^{\dot{\alpha}})$ on $V = 2_{-1} \oplus 2_1$, the holomorphic
$(2,0)$-form simplifies to
\begin{equation}
	\Omega = dz^1 \wedge dz^2 \: .
\end{equation}
This allows us to rewrite the supersymmetry transformation of the one-form ghost
as
\begin{equation}
	\delta_Q C^{(1)} = \iota_v \Omega = v_{+ \dot{\alpha}} dz^{\dot{\alpha}} \: .
\end{equation}
Thus, we again see that the holomorphic components of $C^{(1)}$ cancel with the
diffeomorphism ghost.
\subparagraph{The two-form field $C^{(2)}$}
Let us continue with the supersymmetry transformation of the two-form
\begin{equation} \label{eq:2f susy}
	\delta_Q C^{(2)}_{\mu \nu} = \frac{1}{2} \epsilon_- \Gamma_{\mu \nu} \omega + \epsilon_- \Gamma_{[\mu \rho} \epsilon_- g^\rho_{\nu]} \: .
\end{equation}
The two-form and the graviton decompose into components
\begin{equation}
	\begin{split}
	C^{(2)}_{\mu\nu} &\rightarrow (C^{(2)}_{mn}, C^{(2)}_{m + \dot{\alpha}}, C^{(2)}_{m - \dot{\alpha}}, C^{(2)}_{2} , C^{(2)}_{0} ,C^{(2)}_{(\dot{\alpha} \dot{\beta})} , C^{(2)}_{-2} ) \\
	g_{\mu \nu} &\rightarrow (g_{mn}, g_{m + \dot{\alpha}}, g_{m - \dot{\alpha}}, g_{2 (\dot{\alpha} \dot{\beta})} , g_{(\dot{\alpha} \dot{\beta})}, g_0, g_{-2 (\dot{\alpha} \dot{\beta}) } , h) \: .
	\end{split}
\end{equation}
Consulting the $\Gamma$-matrix decomposition in Equation~\eqref{eq:gamma2
decomp}, we get
\begin{equation}
\label{eq:C2susy}
\begin{split}
\delta_Q C^{(2)}_{mn} &= 0 \\
\delta_Q C^{(2)}_{+ m \dot{\alpha}} &= 0 \\
\delta_Q C^{(2)}_{- m \dot{\alpha}} &= \omega_{m \dot{\alpha}} + g_{+ m\dot{\alpha}} \\
\delta_Q C^{(2)}_{2} &= 0 \\
\delta_Q C^{(2)}_{0} &= \omega_+ \\
\delta_Q C^{(2)}_{(\dot{\alpha} \dot{\beta})} &= g_{2 (\dot{\alpha} \dot{\beta})} \\
\delta_Q C^{(2)}_{-2} &= \omega_- + g_0 \: . 
\end{split}
\end{equation}
Thus we find that the components
\begin{equation}
	C^{(2)}_{- m \dot{\alpha}} \quad C^{(2)}_{0} \quad C^{(2)}_{(\dot{\alpha} \dot{\beta})} \quad C^{(2)}_{-2}
\end{equation}
do not appear in the twisted multiplet, while
\begin{equation}
	C^{(2)}_{mn} \quad C^{(2)}_{+ m \dot{\alpha}} \quad C^{(2)}_{2}
\end{equation}
are in the kernel of $\delta_Q$ and thus, since there are no supersymmetry
transformations that could make these exact, part of the twisted multiplet. Note
again that this matches with the expectation that only $(0,*)$-forms on $M^4$
play a role in the twisted multiplet.

Note that we can rewrite the piece of the supersymmetry
transformation~(\ref{eq:2f susy}) involving the graviton using the holomorphic
(2,0)-form $\Omega$ as
\begin{equation}
	\delta_Q C^{(2)} = \iota_{g^\rho_\nu \partial_\rho} \Omega \wedge dx^\nu \: .
\end{equation}
However, due to the symmetry properties of the graviton, this transformation
alone does not cancel all holomorphic component of the two-form. So one really
needs the supersymmetry ghost to cancel the singlet $C^{(2)}_0$.
\subparagraph{The three-form field $C^{(3)}$}
For the three-form field, we have a supersymmetry transformation of the form
\begin{equation}
	\delta_Q C^{(3)}_{\mu \nu \rho} = \frac{1}{4} \epsilon_- \Gamma_{[\mu \nu} \psi_{\rho]} \: .
\end{equation}
The three-form decomposes into components
\begin{equation}
	C^{(3)}_{\mu \nu \rho} \rightarrow (C^{(3)}_{mnp}, C^{(3)}_{mn+\dot{\alpha}},
	C^{(3)}_{mn-\dot{\alpha}}, C^{(3)}_{m-2}, C^{(3)}_{m0},
	C^{(3)}_{m(\dot{\alpha}\dot{\beta})}).
\end{equation}
To decompose this transformation, we write for the gravitino
\begin{equation}
	\psi^\alpha_\mu = \xi^\alpha \otimes \chi_\mu
\end{equation}
where $\xi^\alpha$ takes values in $S_{11}$ and $\chi_\mu$ in $V_{11}$.
From~(\ref{eq:gamma2 decomp}), we see that $\xi^\alpha$ has to live in
\begin{equation}
	1_{-1} \oplus 1_{1} \oplus (10) 2_0
\end{equation} 
to get a non-zero result. Decomposing $(1_{-1} \oplus 1_{1} \oplus (10) 2_0)
\otimes V_{11}$ into irreducibles, we can identify the decomposed
transformations. The results are listed in
Table~\ref{tab:11DtransformationsDecomp}.
\subparagraph{The supersymmetry ghost $\omega$}
The non-derivative part of the supersymmetry transformation of $\omega_\alpha$
reads
\begin{equation}
\delta_Q \omega_\alpha = \frac{1}{2} (\epsilon \Gamma^{\mu \nu})_{\alpha} (\epsilon \Gamma_{\mu} \psi_{\nu}) \: .
\end{equation}
Again decomposing the gravitino as we did for the three-form field and using the
decomposition~(\ref{eq:gamma decomp}), we find that $\xi^\alpha$ has to take
values in
\begin{equation}
	2_0 \oplus (10) 1_1 \: .
\end{equation}
Tensoring with the vector representation $V_{11}$ and identifying matching
representations gives the result listed below.
\subparagraph{The graviton $g_{\mu \nu}$}
The supersymmetry transformation
\begin{equation}
	\delta_Q g_{\mu \nu} = \frac{1}{2} \epsilon \Gamma_{(\mu} \psi_{\nu )}
\end{equation}
again only allows for $\xi$ to come from $2_0 \oplus (10)1_1$. As before, we
just list the results in Table~\ref{tab:11DtransformationsDecomp}.

In Table~\ref{tab:11DtransformationsDecomp}, we collect all decomposed
non-derivative supersymmetry transformations. Here $M$ is an index for the
14-dimensional representation (01) of $G_2$. It appears in the variation
\begin{equation}
\delta_Q C^{(3)}_{mn -\dot{\alpha}} = \psi_{M \dot{\alpha}} + \psi_{m \dot{\alpha}}
\end{equation}
where the notation describes the decomposition $\wedge^2 (10) \rightarrow (10)
\oplus (01)$ of $G_2$-representations.

\begin{table}[h]
	\caption{Decomposed supersymmetry transformations}
	\begin{center}
		\begin{tabular}{|c|c|}
			\hline
			Operator $\phi$ & Transformation rule $\delta_Q \phi$ \\
			\hline
			$C^{(0)}$  & $0$ \\
			\hline
			$C^{(1)}_m$, $C^{(1)}_{+ \dot{\alpha}}$, $C^{(0)}_{- \dot{\alpha}}$
			&$0$, $0$, $v_{+ \dot{\alpha}}$ \\
			\hline
			$C^{(2)}_{mn}$, $C^{(2)}_{+m \dot{\alpha}}$, $C^{(2)}_{-m
			\dot{\alpha}}$, $C^{(2)}_2$, $C^{(2)}_0$, $C^{(2)}_{(\dot{\alpha}
			\dot{\beta})}$, $C^{(2)}_{-2}$ &$0$,$0$, $\omega_{m \dot{\alpha}} +
			g_{+m \dot{\alpha}}$, $0$, $\omega_+$, $g_{2(\dot{\alpha}
			\dot{\beta})}$, $\omega_- + g_0$ \\
			$v_m$, $v_{+ \dot{\alpha}}$, $v_{- \dot{\alpha}}$ & $\omega_{+ m}$,
			$0$, $\omega_{\dot{\alpha}}$ \\
			\hline
			$\omega_+$, $\omega_-$, $\omega_{\dot{\alpha}}$, $\omega_{-m}$,
			$\omega_{+m}$, $\omega_{m \dot{\alpha}}$ & $0$, $\psi_+$, $0$,
			$\psi_{+m}$, $0$, $\psi_{2 m \dot{\alpha}}$ \\
			$C^{(3)}_{mnp}$, $C^{(3)}_{mn+\dot{\alpha}}$,
			$C^{(3)}_{mn-\dot{\alpha}}$, $C^{(3)}_{m-2}$, $C^{(3)}_{m0}$,
			$C^{(3)}_{m(\dot{\alpha}\dot{\beta})}$ & $0$, $0$, $\psi_{M
			\dot{\alpha}} + \psi_{m \dot{\alpha}}$, $\psi_{m-}$, $\psi_{m+}$,
			$\psi_{m+(\dot{\alpha} \dot{\beta})}$ \\
			 $C^{(3)}_{m2}$, $C^{(3)}_{-\dot{\alpha}}$,
			 $C^{(3)}_{+\dot{\alpha}}$ & $0$, $\psi_{\dot{\alpha}}$,
			 $\psi_{2\dot{\alpha}}$ \\
			$g_{mn}$, $g_{m+\dot{\alpha}}$, $g_{m-\dot{\alpha}}$,
			$g_{2(\dot{\alpha} \dot{\beta})}$, $g_{(\dot{\alpha} \dot{\beta})}$,
			$g_{0}$, $g_{-2(\dot{\alpha} \dot{\beta})}$, $h$ & $\psi_{mn+}$,
			$\psi_{2m \dot{\alpha}}$, $\psi_{m \dot{\alpha}}$, $0$,
			$\psi_{+(\dot{\alpha} \dot{\beta})}$, $\psi_+$,
			$\psi_{-(\dot{\alpha} \dot{\beta})}$, $\psi_+$  \\
			\hline
			$\psi$ & $\delta_Q \psi^\alpha_\mu = (\epsilon M_{\mu \nu}^{\alpha
			\beta} \epsilon) (\psi^{\beta}_{\nu})^{\dagger}$ \\
			\hline
		\end{tabular}
	\end{center}
	\label{tab:11DtransformationsDecomp}
\end{table}

\subsection{Supersymmetry variation of the gravitino}
\label{sec:gravitino}
The non-derivative supersymmetry transformation of the gravitino reads
\begin{equation} \label{eq:gravitino susy}
\delta \psi^\alpha_\mu = (\epsilon M_{\mu \nu}^{\alpha \beta} \epsilon) \psi_{\beta}^{\dagger \nu} \: .
\end{equation}
This transformation reflects the fact that the supersymmetry algebra acts only
up to the equations of motions of the gravitino. Correspondingly, there is a
quadratic term in antifields appearing in the BV action  \cite{Baulieu:1990uv,
Berkovits:2002uc}
\begin{equation}
S^{(2)} \propto (\epsilon M \epsilon) \psi^{\dagger} \psi^{\dagger} \: .
\end{equation}
The transformation~\eqref{eq:gravitino susy} is responsible for the remaining
cancellations between of the gravitino. To argue that indeed the correct
components of $\psi$ cancel, we change our strategy. As the structure of
$M^{\alpha \beta}_{\mu \nu}$ is very complicated, we will not decompose it
directly under $G_2 \times SU(2)$. Instead we give an indirect argument.

For this, recall that~\eqref{eq:gravitino susy} is precisely the term that
corrects for the failure of the linearized supersymmetry transformation to act
strictly. Denoting the linearized part of the supersymmetry transformation by
$\delta_Q^{\mathrm{lin}}$ and the quadratic transformation of the gravitino by
$\delta_Q^{\mathrm{quad}}$, we have
\begin{equation}
\begin{split}
	\{\delta_Q^{\mathrm{lin}} ,\delta_Q^{\mathrm{lin}} \} \psi &= \delta_{\{Q,Q\}}^{\mathrm{lin}} \psi + \delta^{\mathrm{quad}}_Q Q_{BV} \psi^\dagger \\ 
	&= \delta^{\mathrm{quad}}_Q Q_{BV} \psi^\dagger \\
	&= (\epsilon_- M \epsilon_-) \Q_{BV} \psi^\dagger \: ,
\end{split}
\end{equation}
where we have used the fact that $Q$ is square zero in the second equality.

Thus, for $\psi$ outside of the kernel of $Q_{BV}$,
\begin{equation}
	\{\delta_Q^{\mathrm{lin}} ,\delta_Q^{\mathrm{lin}} \} \psi = 0 \implies \delta^{\mathrm{quad}}_Q \psi = 0 \: .
\end{equation}
For such components there can not be any cancellations between $\psi$ and
$\psi^\dagger$. Furthermore this reasoning suggests to view the cancellations
between components of the gravitino and its antifield as a two-step procedure.
First, the linearized transformation identifies a piece of $\psi$ with a
component of $G^{(4)} = dC^{(3)}$. Then we can act with another linearized
transformation to obtain a component of $\psi^\dagger$. Clearly the $U(1)$
charges of components connected in this way satisfy
\begin{equation}
d_{U(1)}(\psi^\dagger) = d_{U(1)}(G^{(4)}) +1 = d_{U(1)}(\psi) +2 \: .
\end{equation}  

Now recall that the linear part of supersymmetry transformations on the
three-form and the gravitino are
\begin{align} 
\delta_Q^{\mathrm{lin}} \psi_{\mu} & = (\Gamma^{\nu \rho \sigma \tau}_{\mu} - 8 \Gamma^{\rho \sigma \tau} \delta_{\mu}^{\nu} )G^{(4)}_{\nu \rho \sigma \tau} \epsilon_- \label{eq:deriv psi} \\ \label{eq:delta C}
\delta_Q^\mathrm{lin} C^{(3)}_{\mu \nu \rho} & = \frac{1}{4} \epsilon_- \Gamma_{[\mu \nu} \psi_{\rho]} \: .
\end{align}
However, from Table~\ref{tab:11DtransformationsDecomp} we know that the
components
\begin{equation} \label{eq:ker of delta}
C^{(3)}_{mnp} \: , C^{(3)}_{mn + \dot{\alpha}} \: ,C^{(3)}_{m 2}
\end{equation}
are in the kernel of $\delta_Q$. Thus the pieces of $\psi$, which are mapped to
the corresponding field strengths by~\eqref{eq:deriv psi} are annihilated by
applying the second linear transformation~\eqref{eq:delta C} and hence satisfy
$\{\delta_Q^\mathrm{lin}, \delta_Q^{\mathrm{lin}}\} \psi = 0$.

With this information, we can analyze the components of the gravitino. In
Table~\ref{tab:twosusys2}, we display the $G_2 \times SU(2)$-equivariant
decomposition of the gravitino, its antifield, and the field strength organized
by $U(1)$ charges. All components of $\psi$ and $\psi^\dagger$ that form trivial
pairs with other fields according to Table~\ref{tab:11DtransformationsDecomp}
are indicated with an arrow.
\begin{table}[htp]
	\tiny
	\caption{Decomposition of the non-linear gravitino supersymmetry variation}
	\begin{center}
		\begin{tabular}{|c|c|c|c|c|c|}
			\hline
			Field & 2 & 1 & 0 & -1 & -2 \\
			\hline
			$\psi$  & $\cancelto{}{(00)(2)} \oplus \cancelto{}{(10)(2)}$  &  $
			\circled{(10)(1) \oplus (01)(1)}$ &  $\cancelto{}{(00)(2)} \oplus
			\cancelto{}{(10)(2)} \oplus \cancelto{}{(01)(2)}$ &
			$\cancelto{}{(10)(1)} \oplus \greencircled{(01)(1)}$	 &
			$\greencircled{(00)(2) \oplus (10)(2)}$ \\
			&   & $\cancelto{}{(00)(1)} \oplus \cancelto{}{(10)(1)} \oplus
			\cancelto{}{(20)(1)}$ & $\circled{(00)(2) \oplus (10)(2) \oplus
			(20)(2)}$ & $\circled{(00)(1) \oplus (10)(1) \oplus (20)(1)}$ & \\
			&   & $\cancelto{}{(00)(3)} \oplus \cancelto{}{(10)(3)}$ &
			$\cancelto{}{(10)(2)}$ & $\cancelto{}{(00)(3)} \oplus
			\greencircled{(10)(3)}$ & \\			 		 
			& & $\cancelto{}{(00)(1)} \oplus \cancelto{}{(10)(1)}$ &
			$\greencircled{(00)(2) \oplus (10)(2)} $ & $\greencircled{(00)(1)
			\oplus (10)(1)}$ &  \\
			\hline
			$G^{(4)}$ & $(10)(1) \oplus (01)(1)$ & $ (10)(2) \oplus (20)(2)$  &
			$(00)(1) \oplus (20)(1) \oplus (10)(1)$ & $(20)(2) \oplus (10)(2)$
			& $(10)(1) \oplus (01)(1)$  \\
			&  & $\greencircled{(00)(2) (10)(2)}$  & $\greencircled{(00)(1)
				\oplus (10)(1) \oplus (10)(3)}$ &$\greencircled{(00)(2) \oplus
				(10)(2)}$ &  \\
			&  &  & $\greencircled{(01)(1)} \oplus (01)(3)$ & &  \\
			\hline
			$\psi^{\dagger}$  & $\greencircled{(00)(2) \oplus (10)(2)}$  &
			$\cancelto{}{(10)(1)} \oplus \greencircled{(01)(1)}$ &
			$\cancelto{}{(00)(2)} \oplus \cancelto{}{(10)(2)} \oplus
			\cancelto{}{(01)(2)}$ & $\circled{(10)(1) \oplus (01)(1)}$	 &
			$\cancelto{}{(00)(2)} \oplus \cancelto{}{(10)(2)}$ \\
			&   & $\circled{(00)(1) \oplus (10)(1) \oplus (20)(1)}$ &
			$\circled{(00)(2) \oplus (10)(2) \oplus (20)(2)}$ & $
			\cancelto{}{(00)(1)} \oplus \cancelto{}{(10)(1)} \oplus
			\cancelto{}{(20)(1)}$ & \\
			&   & $\cancelto{}{(00)(3)} \oplus \greencircled{(10)(3)}$ &
			$\cancelto{}{(10)(2)}$ & $\cancelto{}{(00)(3)} \oplus
			\cancelto{}{(10)(3)}$ & \\			 		 
			& & $\greencircled{(00)(1) \oplus (10)(1)}$ & $\greencircled{(00)(2)
			\oplus (10)(2)} $ & $\cancelto{}{(00)(1)} \oplus
			\cancelto{}{(10)(1)}$ &  \\
			\hline
		\end{tabular}
	\end{center}
\label{tab:twosusys2}
\end{table}

We immediately see that the components of $\psi$ with $U(1)$ charge $1$ cannot
be canceled and thus are part of the twisted multiplet. We circle these
components in blue.

Furthermore, we can take a look at the remaining components of $\psi$ with
$U(1)$ charge $-1$. There we have a representation
\begin{equation}
	(00)(1) \oplus (10)(1) \oplus(20)(1) \cong \Omega^4(M^7) \otimes \Omega^{0,0}(M^4) \: ,
\end{equation}
which maps under $\delta_Q$ to $d_{M^7} C^{(3)}_{mnp}$. This means the
corresponding components are part of the twisted multiplet. With similar
reasoning the components
\begin{equation}
	(00)(2) \oplus (10)(2) \oplus (20)(2) \cong \Omega^3(M^7) \otimes \Omega^{0,1}(M^4)
\end{equation}
with $U(1)$ charge zero transform to the field strength of $d_{M^7} C^{(2)}_{mn
+ \dot{\alpha}}$ and $\bar{\partial} C^{(3)}_{mnp}$ under $\delta_Q$ and hence
are also part of the twisted multiplet.

On the other hand, we see that different pieces of the gravitino are mapped to
components of the field strength which are not part of the kernel of $\delta_Q$.
These than can have $\{\delta_Q^\mathrm{lin}, \delta_Q^{\mathrm{lin}}\} \psi
\neq 0$, such that a cancellation is possible. In Table~\ref{tab:twosusys2} we
indicate such components, the corresponding intermediate components of the field
strength and the respective partners from $\psi^\dagger$ with green rectangles.

Nevertheless one has to remain careful. As we explained above, these arguments
only hold outside of the kernel of $Q_{BV}$. For $U(1)$ charge zero, there is a
component $(00)(2)$ boxed in green. This can be viewed as a differential form
\begin{equation}
	(00)(2) \cong \Omega_1^3(M^7) \otimes \Omega^{0,1}(M^4) \subset \Omega^3(M^7) \otimes \Omega^{0,1}(M^4) \: .
\end{equation}
The corresponding field strength, however, does come from $C^{(3)}_{mn
+\dot{\alpha}}$ which is in the kernel of $\delta_Q$. This is not a
contradiction, since the corresponding representation $(00)(2)$ is in the kernel
of $Q_{BV}$. The trivial representation $(00) \subset \Omega^3(M^7)$ corresponds
to a covariantly constant spinor inside the tensor product $(TM^7)^\C \otimes
SM^7$~\cite{Guio:2017zfn}, which is a zero-mode for the BV operator $Q_{BV}$
which acts as the Rarita--Schwinger operator. This means that the above argument
does not apply here, in the light of the results so far and the conjecture, we
nevertheless expect this component to cancel. An explicit investigation using a
decomposition of the tensor $M^{\alpha \beta}_{\mu \nu}$ would still be
interesting.

\subsection{Summary of cancellations}
We summarize the cancellations obtained in the previous sections in
Table~\ref{tab:cancellations}.  The fields that do not form trivial pairs are
circled in blue. They form the multiplet $\cA \in \Omega^{\bullet}(M^7) \otimes
\Omega^{0,\bullet}(M^4)$ and appear in Table \ref{tab:twistedfields}. The
bi-directional strike-through arrows indicate cancellations that occur between
$\psi$ and its anti-field $\psi^{\dagger}$ found in Section~\ref{sec:gravitino}.

Special care should be taken for the variations of the components of $C^{(2)}$
that cancel with a linear combination of components of the graviton and
supersymmetry ghost
\begin{align}
\delta_Q C^{(2)}_{- m \dot{\alpha}} &= \omega_{m \dot{\alpha}} + g_{+ m\dot{\alpha}} \\
\delta_Q C^{(2)}_{-2} &= \omega_- + g_0 \:  
\end{align}
that occur in Equation~\eqref{eq:C2susy}. A subsequent variation yields
\begin{align}
\delta_Q \omega_{m \dot{\alpha}} & = - \delta_Q g_{+ m\dot{\alpha}} = \psi_{2 m \dot{\alpha}} \\
\delta_Q \omega_{-} & = -\delta_Q g_0 = \psi_{+}
\end{align}
which is consistent with $\delta_Q^2 C^{(2)}= 0$. These extra cancellations are
indicated by the strike-through arrows with labels $x$ and $y.$

\begin{table}[htp]
\tiny

	\caption{Cancellations of fields under $Q$.   Fields are decomposed into
	$G_2 \times SU(2)_{-} \times U(1)_L$-representations.}
	\begin{center}
		\begin{tabular}{|c|c|c|c|c|c|c|}

			\hline
			Field & $Spin(11)$ & 2 & 1 & 0 & -1 & -2\\
			\hline
			$C^{(0)}$ & $(00000)$ & & & \circled{(00)(1)}&  & \\
			\hline
			$C^{(1)}$ & $(10000)$ & & \circled{(00)(2)} & \circled{(10)(1)}&
			$\cancelto{a}{(00)(2)}$ &  \\
			\hline
			$C^{(2)}$ & $(01000)$ & \circled{(00)(1)} & \circled{(10)(2)} &
			$\cancelto{b}{(00)(1)} \oplus \cancelto{c}{(00)(3)} \oplus
			\circled{(10)(1) \oplus (01)(1)}$ & $\cancelto{\; \; d,x}{(10)(2)}$
			& $\cancelto{\; \; d,y}{(00)(1)}$ \\
			\hline
			$v$ & $(10000)$ & & $\cancelto{a}{(00)(2)}$ &
			$\cancelto{e}{(10)(1)}$ & $\cancelto{f}{(00)(2)}$ &  \\
			\hline
			$\omega$ & $(00001)$ & & $\cancelto{b}{(00)(1)} \oplus
			\cancelto{e}{(10)(1)}$ & $\cancelto{f}{(00)(2)} \oplus \cancelto{\;
			\; g, x}{(10)(2)} $ & $\cancelto{\; \; h, y}{(00)(1)} \oplus
			\cancelto{h}{(10)(1)}$ &  \\
			\hline
			$C^{(3)}$ & $(00100)$ & \circled{(10)(1)} & $\cancelto{i}{(00)(2)}
				\oplus \circled{(10)(2) \oplus (01)(2)}$ & $ \circled{(00)(1)
				\oplus (10)(1) \oplus (20)(1)}$ & $\cancelto{k}{(00)(2)} \oplus
				\cancelto{k}{(10)(2)} \oplus \cancelto{k}{(01)(2)}$ &
				$\cancelto{l}{(10)(1)}$ \\
			 &  &  &  &$\cancelto{j}{(10)(1)} \oplus \cancelto{j}{(10)(3)}$ &  &
			 \\

			\hline
			$g$ & $(20000)$ & $\cancelto{c}{(00)(3)}$& $\cancelto{d}{(10)(2)}$ &
			$\cancelto{d}{(00)(1)} \oplus \cancelto{m}{(00)(3)} \oplus
			\cancelto{m}{(20)(1)}$& $\cancelto{o}{(10)(2)}$  &
			$\cancelto{p}{(00)(3)}$\\
			 & $(00000)$ & &  & $\cancelto{n}{(00)(1)}$ &   & \\

			\hline
			$\psi$ & $(10001)$ & $\cancelto{i}{(00)(2)} \oplus
			 \cancelto{g}{(10)(2)}$  &  $ \circled{(10)(1) \oplus (01)(1)}$ &
			 $\cancelto{k}{(00)(2) }\oplus \cancelto{k}{(10)(2)} \oplus
			 \cancelto{k}{(01)(2)}$ & $\cancelto{l}{(10)(1)} \oplus
			 \xcancelto{q}{(01)(1)}$	  & $\xcancelto{q}{(00)(2)} \oplus
			 \xcancelto{q}{(10)(2)}$ \\
			 &  &   & $\cancelto{h}{(00)(1)} \oplus \cancelto{h}{(10)(1)} \oplus
			 \cancelto{m}{(20)(1)}$ & $\circled{(00)(2) \oplus (10)(2) \oplus
			 (20)(2)}$ & $ \circled{(00)(1) \oplus (10)(1) \oplus (20)(1)}$ & \\
			 &  &   & $\cancelto{m}{(00)(3)} \oplus \cancelto{j}{(10)(3)}$ &
			 $\cancelto{o}{(10)(2)}$ & $\cancelto{p}{(00)(3)} \oplus
			 \xcancelto{q}{(10)(3)}$ & \\			 		 
			 & $(00001)$ & & $\cancelto{n}{(00)(1)} \oplus
			 \cancelto{j}{(10)(1)}$ & $\xcancelto{q}{(00)(2)} \oplus
			 \xcancelto{q}{(10)(2)} $ & $\xcancelto{q}{(00)(1)} \oplus
			 \xcancelto{q}{(10)(1)}$ &  \\
			\hline
		\end{tabular}
	\end{center}
	\label{tab:cancellations}
\end{table}

\subsection{The twisted differential}
Recall that the BV differential of the twisted theory is the sum of two terms
\begin{equation}
	Q_{BV}^{\text{tw}} = Q_{BV} + \delta_Q \: .
\end{equation}
We already examined how the non-derivative part of $\delta_Q$ leads to the
formation of various trivial pairs; now we turn towards the parts containing
derivatives in order to see how they act on the twisted multiplet.

The BV operator $Q_{BV}^{\mathrm{tw}}$ is dual to a differential
$D^{\mathrm{tw}}$ acting on the fields of the twisted multiplet. We already know
that $D$ acts as the de Rham differential on the three-form ghost system. Under
$G_2 \times SU(2)$ the de Rham differential decomposes
\begin{equation}
	d = d_{M^7} + \bar{\partial}_{M^4} + \partial_{M^4} \: .
\end{equation} 
As only $(0,*)$-forms are part of the twisted multiplet, this restricts to
\begin{equation}
	d_{M^7} + \bar{\partial}_{M^4} \: .
\end{equation}
In addition, $D$ acts on the gravitino by the Rarita--Schwinger equation.
Identifying the gravitino as a spinor valued one-form, $\psi \in \Omega^1(M)
\otimes S_{11}$, the Rarita--Schwinger operator can be understood as a
composition of the exterior differential and Clifford
multiplication~\cite{Homma:2018hmn}. From this one can see that it also acts by
$d_{M^7} + \bar{\partial}_{M^4}$ on the relevant pieces of the gravitino.

Finally, there is a contribution to $D^{\mathrm{tw}}$ coming from the
supersymmetry transformation~\eqref{eq:delta C}. This transformation also acts
by $d_{M^7} + \bar{\partial}_{M^4}$ and provides the missing differential
between $C^{(3)}$ and $\psi$.

In summary, the twisted multiplet can thus be described by the cochain complex
\begin{equation}
	\left( \Omega^{\bullet}(M^7) \otimes \Omega^{0,\bullet}(M^4) \: , \: D^{\mathrm{tw}} = d_{M^7} + \bar{\partial}_{M^4} \right) \: , 
\end{equation}
as conjectured by Costello.

Interestingly, the form of the differential can also be deduced directly from
the explicit formulas in the pure spinor formalism. Recall that $\cD_1$ acts on
the representatives by
\begin{equation}
	\cD_1 = (\lambda \Gamma^\mu \theta) \partial_\mu \: ,
\end{equation}
and that the one-form was represented by the cohomology classes $C^{(1)}_\mu
(\lambda \Gamma^\mu \theta)$. As we already know that the twisted multiplet
forms the exterior algebra $\Omega^{\bullet}(M^7) \otimes
\Omega^{0,\bullet}(M^4)$, we see that $\cD_1$ simply acts by taking derivatives
and wedging with the corresponding component of the one-form, i.e. precisely by
$d_{M^7} + \bar{\partial}_{M^4}$.

In addition the derivative part of the supersymmetry transformation acts by 
\begin{equation}
	Q_{\partial_x} = (\epsilon_- \Gamma^\mu \theta) \partial_\mu \: .
\end{equation}
From the Gamma matrix decomposition~(\ref{eq:eps decomp}), we see
\begin{equation}
	(\epsilon_- \Gamma^\mu \theta) \in 2_{-1} \oplus (10) \: .
\end{equation}
Identifying the corresponding components with $d\bar{z}^{\dot{\alpha}}$ and
$dx^m,$ we once again see that $Q_{\partial_x}$ acts as desired.

A more roundabout way of understanding the appearance of the de Rham
differential is as follows. Recall that the gravitino field on $M^7$ can be
organized into $\Omega^2 \oplus \Omega^3$ when $M^7$ has $G_2$ holonomy. Since
there are $b^2(M^7) + b^3(M^7)$ zero modes of the gravitino on $M^7$
\cite{Font:2010sj, Guio:2017zfn, MR1129331, Homma:2018hmn}, we see that the BV
differential acts by the de Rham differential
\begin{equation}
d_{dR}: \Omega^2 \oplus \Omega^3 \rightarrow \Omega^3 \oplus \Omega^4.
\end{equation}

This is similar to the holomorphic twist of ten-dimensional abelian super
Yang--Mills theory on $\C^5$. In that case, the analogous BV differential
between the gaugino and its antifield expresses the Dirac equation.  The
relevant part of the differential in the twisted theory is
\begin{equation}
Q_{BV} (\lambda^{mn})^{\dagger} = i \epsilon^{mnpqr} \delbar_p \lambda_{qr},
\end{equation}
and only involves the Dolbeault operator on $\Omega^{0,\bullet}(\C^5)$.

\section{Conclusions and future directions}
\label{sec:conclusions}
Eleven-dimensional supergravity in the pure spinor formalism incorporates both
the three-form and super-vielbein multiplets. We have seen how the twist of the
three-form multiplet is given by a differential form
\begin{equation}
\cA \in \Omega^{\bullet}(M^7) \otimes \Omega^{0,\bullet}(M^4).
\end{equation}
The super-vielbein multiplet has the graviton, gravitino, and 4-form field
strength $G^{(4)}$ as its physical fields. It is used in the traditional
superspace formulation of supergravity. It is natural to expect that the twisted
fields of the super-vielbein multiplet organize into a differential form
\begin{equation}
\partial \cA \in \Omega^{\bullet}(M^7) \otimes \Omega^{1,\bullet}(M^4),
\end{equation}
with leading component $v_{+ \dot{\alpha}}$ from the diffeomorphism ghost. In
future work \cite{rfsw2021}, we plan to directly twist Cederwall's pure spinor
action \cite{Cederwall:2010tn} and compare to Costello's conjectural action
\cite{Costello:2016nkh, Raghavendran:2019zdq} for the twisted theory.

The conjectural twist of type IIB supergravity was developed by Costello and Li
to give a precise formulation of a sub-sector of AdS/CFT with rigorously defined
mathematical objects \cite{Costello:2015xsa,Costello:2016mgj}. We hope that a
similar approach can be used to derive the holomorphic twist of M-theory and
Costello--Li's conjectural form of the twist of type IIB supergravity as a BCOV
theory using the presymplectic BV formalism of \cite{Saberi:2020pmw}.

The AdS/CFT conjecture is a holographic duality between string and M-theory on
anti-deSitter spaces and gauge theories.  In a particular limit it relates
weakly coupled type IIB supergravity on products of five-dimensional AdS space
$AdS_5$ with arbitrary Sasaki-Einstein manifolds $SE^5$ to four-dimensional
supersymmetric gauge theories. A different form of the conjecture relates the
weak coupling limit of M-theory on the products $AdS_4 \times SE^7$ to
three-dimensional supersymmetric gauge theories. The cone over the
Sasaki--Einstein manifold is a local Calabi--Yau manifold. One corollary of the
conjecture is the equivalence of the superconformal index
\cite{Romelsberger:2005eg, Kinney:2005ej} under gauge-gravity duality. The
gravity superconformal index was computed in terms of holomorphic invariants of
the Calabi--Yau manifold in \cite{Eager:2012hx, Eager:2013mua}. The
corresponding field theory index was later shown to be most directly computed in
the holomorphic twist \cite{Eager:2018oww, Saberi:2019fkq}. Thus a full
derivation of the holomorphic twist of type IIB supergravity and
eleven--dimensional supergravity should reproduce the index calculations of
\cite{Eager:2012hx, Eager:2013mua}. This would serve as a natural bridge between
physical and mathematical approaches to holography.

We hope that a further twist of the one considered in this paper can be used to
derive twisted M-theory in the $\Omega$-background \cite{Costello:2016nkh}
following \cite{Oh:2019bgz}. This could provide a physical origin for the
applications in \cite{Gaiotto:2019wcc, Oh:2021bwi} by coupling a twisted
M5-brane \cite{Saberi:2020pmw} to twisted M-theory. Finally, we hope that
twisted M-theory can shed new light on topological M-theory
\cite{Hitchin:2000jd, Gerasimov:2004yx, Dijkgraaf:2004te, Grassi:2004xr,
Becker:2016edk}, which is believed to unify the K\"ahler
\cite{Bershadsky:1994sr}  and Kodaira--Spencer theories of topological gravity.

\subparagraph{Acknowledgements}

R.E. and F.H. would like to thank I. Saberi, J. Walcher, and B. Williams for
collaboration on related projects and K. Costello, W. Linch, and S. Raghavendran
for discussions.

\bibliographystyle{ytphys}
\bibliography{twistedMtheory}
\end{document}